\documentclass[12pt,preprint]{aastex}

\newcommand{\mstar}{M_{\star}}
\newcommand{\lstar}{L_{\star}}

\newcommand   {\av}    {\mbox{${\rm A_v}$}}

\renewcommand {\deg}   {\mbox{$^\circ$}}

\newcommand   {\kms}   {\mbox{km\,s$^{-1}$}}
\renewcommand {\ga}    {\mbox{\rlap{\hbox{\lower5pt\hbox{$\sim$}}}\hbox{$>$}}}
\renewcommand {\la}    {\mbox{\rlap{\hbox{\lower5pt\hbox{$\sim$}}}\hbox{$<$}}}

\begin{document}

%\author{ }
%\email{ }
%\altaffiltext{1}{ }

%\begin{abstract}
%\vspace {5pt}
%\end{abstract}

%\keywords{ }
%\documentclass[12pt,preprint]{aastex}
%% manuscript produces a one-column, double-spaced document:
%% \documentclass[manuscript]{aastex}
%% preprint2 produces a double-column, single-spaced document:
%\documentclass[preprint2]{aastex}
%% Sometimes a paper's abstract is too long to fit on the
%% title page in preprint2 mode. When that is the case,
%% use the longabstract style option.
%% \documentclass[preprint2,longabstract]{aastex}
\def\kms {\hbox{km{\hskip0.1em}s$^{-1}$}} % km/s
\def\msol{\hbox{$\hbox{M}_\odot$}}
\def\lsol{\hbox{$\hbox{L}_\odot$}}
\def\kms{km s$^{-1}$}
\def\Blos{B$_{\rm los}$}
\def\etal   {{\it et al. }}                     % et al
\def\psec           {$.\negthinspace^{s}$}
\def\pasec          {$.\negthinspace^{\prime\prime}$}
\def\pdeg           {$.\kern-.25em ^{^\circ}$}
\def\degree{\ifmmode{^\circ} \else{$^\circ$}\fi}
\def\ee #1 {\times 10^{#1}}          % \ee p       10^p   
\def\ut #1 #2 { \, \textrm{#1}^{#2}} % \ut unit p  unit^p 
\def\u #1 { \, \textrm{#1}}          % \u unit     unit
\def\nH {n_\mathrm{H}}

\def\ddeg   {\hbox{$.\!\!^\circ$}}              % Degrees over dot
\def\deg    {$^{\circ}$}                        % Degrees symbol
\def\le     {$\leq$}                            % <=
\def\sec    {$^{\rm s}$}                        % Second of time
\def\msol   {\hbox{$M_\odot$}}                  % Solar mass
\def\i      {\hbox{\it I}}                      % italic I
\def\v      {\hbox{\it V}}                      % italic V
\def\dasec  {\hbox{$.\!\!^{\prime\prime}$}}     % Arcseconds over dot
\def\asec   {$^{\prime\prime}$}                 % Arcseconds symbol
\def\dasec  {\hbox{$.\!\!^{\prime\prime}$}}     % Arcseconds over dot
\def\dsec   {\hbox{$.\!\!^{\rm s}$}}            % Second over dot
\def\min    {$^{\rm m}$}                        % Minutes of time
\def\hour   {$^{\rm h}$}                        % Hours of time
\def\amin   {$^{\prime}$}                       % Arcminutes symbol
\def\lsol{\, \hbox{$\hbox{L}_\odot$}}
\def\sec    {$^{\rm s}$}                        % Second of time     
\def\etal   {{\it et al. }}                     % \etal

\def\xbar   {\hbox{$\overline{\rm x}$}}         % bar over x

%\slugcomment{Draft 2}
\shorttitle{}
\shortauthors{}

\title{Signatures of  Young   Star Formation Activity\\
Within Two Parsecs of  Sgr A*}
\author{F. Yusef-Zadeh$^{1}$,
M. Wardle$^2$,  M. Sewilo$^3$, 
D. A. Roberts$^1$, I. Smith$^2$, 
R. Arendt$^4$, W. Cotton$^5$, J. Lacy$^6$, 
S. Martin$^7$,  
M. W. Pound$^8$, M. Rickett$^1$ \&  M. Royster$^1$} 
\affil{$^1$Department of Physics and Astronomy and CIERA,
Northwestern University, Evanston, IL 60208}
\affil{$^2$Department of Physics and Astronomy, and Research Center for Astronomy, 
Astrophysics \& Astrophotonics, Macquarie University, Sydney NSW 2109, Australia}
\affil{$^3$Space Science Institute, 4750 Walnut St. Suite 205, Boulder, CO 80301, USA}
\affil{$^4$CRESST/UMBC/NASA GSFC, Code 665, Greenbelt, MD 20771}
\affil{$^5$National Radio Astronomy Observatory,  Charlottesville, VA 22903}
\affil{$^6$Department of Astronomy, University of Texas, Austin, TX 78712}
\affil{$^7$Institut de Radio Astronomie Millim\'etrique, 300 rue de la Piscine, Dom. Univ., 38406 St Martin d'H\`eres, France}
\affil{$^8$Department of Astronomy, University of Maryland, College Park, 
MD  20742}
%\affil{$^9$Department of Astronomy, University of Wisconsin, 475 North Charter Street, Madison, WI 53706
%Department of Astronomy, University of Wisconsin}

%\date{December 5, 2005}
\vfill\eject
%S. Mart\'in^7$,  

\begin{abstract} 

We present radio and infrared observations indicating on-going star formation activity inside the
$\sim2-5$ pc circumnuclear ring at the Galactic center. Collectively these measurements suggest a
continued disk-based mode of on-going star formation has taken place near Sgr A* over the last few million
years. First, VLA observations with spatial resolution 2.17$''\times0.81''$ reveal 13 water masers,
several of which have multiple velocity components. The presence of interstellar water masers suggests gas        
densities that are sufficient for self-gravity to overcome the tidal shear of the 4$\times10^6$ \msol\,
black hole. Second, SED modeling of stellar sources indicate massive YSO candidates interior to the      
molecular ring, supporting in-situ star formation near Sgr A* and appear to show a distribution similar to
that of the counter-rotating disks of $\sim$100 OB stars orbiting Sgr A*. Some YSO candidates (e.g.,
IRS~5) have bow shock structures suggesting that they have have gaseous disks that are phototoevaporated
and photoionized by the strong radiation field.  Third, we detect clumps of SiO (2-1) and (5-4) line
emission in the ring based on CARMA and SMA observations.  The FWHM and                             
luminosity of the SiO emission is consistent with shocked protostellar outflows.  Fourth, two linear              
ionized features with an extent of $\sim0.8$ pc show blue and redshifted velocities between $+50$ and   
$-40$ \kms, suggesting protostellar jet driven outflows with mass loss rates of $\sim5\times10^{-5}$  
\msol\, yr$^{-1}$. Finally, we present the imprint of radio dark clouds at 44 GHz, representing a      
reservoir of molecular gas that feeds star formation activity close to Sgr A*.
\end{abstract}

%We first show  H$_2$O maser observations toward the 2-5 pc circumnuclear molecular ring orbiting Sgr A* 
%This can be  water 
%masers suggest that a small fraction of the gas must have  high density $\sim10^{8}$ cm$^{-3}$ to be  tidally stable.
%A velocity of $\sim-100$ \kms is noted at the western tip of the ridge. 
%The SiO (5-4) clumps with broad and narrow 
%linewidths 

\keywords{Galaxy: center - clouds - ISM: general - ISM - radio continuum - stars: protostars}

\section{Introduction}

A critical question regarding star formation activity near supermassive black holes (SMBHs) is whether tidal shear in the 
vicinity of SMBHs is able to completely suppress star formation or whether it enables  a disk-based mode of star formation, 
entirely distinct from the standard cloud-based mode observed in the Galactic disk. 
The molecular ring orbiting the 
4$\times10^6$~\msol\, black hole Sgr A* in the inner few parsecs of the Galactic center 
is an excellent testing ground to 
examine how star formation proceeds in the extreme tidally sheared environment of  a supermassive black hole. 

%The key in our understanding of star formation comes from 

A fragile and kinematically disturbed circumnuclear molecular ring 
(CMR), also known as the  circumnuclear molecular disk (CND),  orbits Sgr A* with a velocity of 110 km s$^{-1}$  
and is about  2-7 pc in extent 
(Jackson et al. 1993; 
Christopher et 
al.  2005; Montero-Casta\~no et al. 2010; Etxaluze et al. 2011; Goicoechea et al. 2013). 
The Roche number density required for self-gravity to overcome 
tidal shear is n$\sim2\times10^8$ cm$^{-3}$ at 1 pc from Sgr~A* and is substantially higher at sub-pc distances (n $\propto\, 
r^{-3}$). HCN (1-0) observations (Christopher et al. 2005) infer that the clumps in the CMR have densities greater than the 
critical Roche density and that star formation can take place in the molecular ring.  
However, the implied mass of the CMR exceeds 10$^5$ \msol\,  is inconsistent 
with far-infrared (FIR) dust emission and the lack of dynamical effects  attributed to 
gravitational potential of the ring (Genzel, Eisenhauer \& Gillissen 2010).
Indeed, recent analysis of molecular 
line data suggests that the density in the ring has been overestimated by one to two orders of magnitude 
(Requena-Torres et al. 2012; Mills et al. 2013;
Smith and Wardle 2014) 
implying that 
the gas is generally tidally unstable against self-gravity. Nevertheless, 
star formation has taken place in this region in the last few million years, 
 as evidenced by a population of more than  100 OB stars distributed 
in two disks orbiting clockwise and counterclockwise within 10$''$ (0.4 pc) of Sgr A* 
(Beloborodov et al. 2006; Tanner et al. 2005, 
Paumard et al. 2006; Lu et al. 2009). 
Theoretical models have been proposed to explain  star formation  in this environment  
(Nayakshin et al. 2007; Wardle \& Yusef-Zadeh 2008, 2012, 2014; Bonnel \& Rice 2008; 
Mapelli et al. 2012; Mapelli, Gualandris \& Hayfield 2013; Jalali et al. 2014).

Motivated by the discovery of young stars near Sgr A*, 
we have previously  searched for signs of 
on-going star formation on different scales. 
On subpc scales, ALMA observations of the interior of the molecular ring detected SiO (5-4)
emission from 11 clumps (Yusef-Zadeh et al. 2013).  The full linewidths as a function of the luminosity of SiO (5-4) line 
emission are similar to those of low and high mass protostellar systems in the Galactic disk (Gibb  \etal  2004 and 2007).
 This led to the suggestion 
that SiO emission clumps trace protostellar outflows from young stellar objects (YSOs). 
This interpretation was supported by spectral energy distribution (SED) modeling of stellar sources using 
{\it Spitzer} 
Space Telescope ({\it Spitzer}) 
and the Two Micron All Sky Survey (2MASS) data,
which identified two YSO candidates near SiO clumps (Yusef-Zadeh et al. 2013).

%The large linewidths of some of the SiO clumps showed  that 
%they are  not   gravitationally bound. This property of the clumps was used 
%as an  argument for outflows.  

In a  second study, we  detected seven collisionally excited $7_0-6_1 A^+$ methanol masers (44.0694 GHz) and water masers (22.23419 
GHz) within 2 pc of Sgr A* with the Green Bank Telescope (GBT) (Yusef-Zadeh et al. 2008, hereafter YBWR08). The association of 
CH$_3$OH and H$_2$O masers  with molecular clumps in the  ring suggested that 
 hot cores and YSO 
outflows may be present. 
However, 
 interstellar and stellar 
water masers can not be distinguished in low spatial resolution data, so  the suggestion that   
water masers trace  ongoing star formation is tentative.

Here we present five different studies supporting  star formation activity within the inner 2 pc of 
Sgr A*. After describing  the multi-wavelength  observations in $\S2\,$, we  
first focus in $\S3.1\,$
  on high resolution observations 
of  interstellar  22 GHz water maser line 
emission. 
Second, we  concentrate in $\S3.2\,$ 
on SED fitting of the YSO models to infrared excess stellar sources.
We discuss
individual YSO candidates and note their similar distribution to  the counter-rotating stellar disks 
orbiting Sgr A*. 
Third,  we present  in $\S3.3\,$
SiO (2-1) and SiO (5-4) line emission from the inner $3'$ of the CMR. 
Fourth, we present kinematics of an ionized linear feature in $\S3.4$, 
which we  interpret as a  protostellar jet from a YSO. 
Finally, in $\S3.5\,$,  we argue  that as  molecular clouds
are bathed in the intense  radiation  field of the Galactic center, they manifest 
regions of suppressed 
radio continuum emission (Yusef-Zadeh 2012).  Identification of 
such ``radio dark clouds'' implies that there is a supply of dense gas
feeding star formation activity near Sgr A*.

%In addition, these measurements supports the idea that the CMR is young and is possibly captured in the last 
%$\sim10^5$ years. 

% and provide LVG analysis of 
%SiO clumps. 

\section{Observations and Data Reductions}

In the following sections, we briefly describe the data used for our analysis. The water maser  data are 
presented here for the first time; the rest of the data have already been either completely or partly 
published in the past (see references below).

\subsection{Present Data}

\subsubsection{H$_2$O Line at 22 GHz}

Water maser observations were carried out with
the Karl G. Jansky Very Large Array
(VLA)\footnote{Karl G. Jansky Very Large Array (VLA) of the National Radio 
Astronomy Observatory is a facility of the National
Science Foundation, operated under a cooperative agreement by Associated Universities, Inc.}
in its B-configuration on February 2, 2012.  The observations 
used the new correlator to sample 256 channels at 62.5 kHz (0.84 km s$^{-1}$).  
During the hour-long observations, we 
alternated between two pointing positions at 
$\alpha, \delta\, (J2000)=17^h 45^m 43^s, -29^{\circ} 00' 00''$
and $17^h 45^m 39^s, -29^{\circ} 01' 00''$. 
3C286 was used as a primary flux calibrator and NRAO530 as the complex gain and bandpass calibrator. 
The data were calibrated in the usual manner using Astronomical Image Processing System (AIPS). 
 Data for both pointings were phase self-calibrated using Sgr 
A* as a point source model. Continuum subtraction in the visibility plane was carried out using the UVLSF task of 
AIPS before mosaic images were constructed with a spatial and spectral  
resolutions of 2.17$''\times0.81''$ (PA=$-10.9^{\circ}$) and 0.9 \kms, respectively. 
The rms noise  in each channel map is  about 2.8 mJy beam$^{-1}$.

\subsection{Published Data}

\subsubsection{SiO (2-1) and (5-4) Lines}

The SiO (2-1) line data were taken with
the Combined Array for Research in Millimeter-wave Astronomy 
(CARMA)\footnote{Ongoing CARMA development and operations are supported by the National 
Science Foundation and the CARMA partner universities.}
during the 2009 and 2010 observing
seasons in the D and C array configurations.  The array consisted of
six 10.4m antennas and nine 6.1m antennas and the maps were made on a
127-point hexagonal mosaic covering 7.1 arcminutes, Nyquist-sampling the 10.4m antenna primary
beam.  The spatial resolution and          
spectral resolutions of the final maps are $6.8\arcsec\times 3.7\arcsec$ (PA=1$^{\circ}$.2)
and 6.74 \kms, respectively, covering a total Local Standard of Rest (LSR) 
velocity range from $-180$  to 180 \kms.

We also used the SiO (5-4) line data taken with the Sub-Millimeter Array (SMA). 
Details of these observations  are very similar to the data presented 
 in Martin et al. (2012) except  that 
an additional track of data  was added using the SMA in its compact 
configuration. 
After applying  uniform weighting to  
{\it uv} data, 
the CMR was mapped in its entirety in the SiO (5-4) line with a spatial and spectral resolution of 
$3.6''\times2.4''$ 
and 2 \kms\, respectively,  
covering  a total velocity range  between --140 and 140 \kms.

\subsubsection{[Ne\,{\sc ii}] Line \& 12.8$\mu$m Continuum}

Details of the [Ne\,{\sc ii}]
line  observations are given in Yusef-Zadeh et al. (2010) and Irons et al. (2012). 
Briefly,  
Sgr A West  was observed in the [Ne II] (12.8$\mu$m) line with the high-resolution mid-infrared
spectrograph TEXES on the NASA IRTF on 2009 June 3 and 2010 May 30 (UT). TEXES (Lacy et al. 2002) 
used  high-resolution, cross-dispersed, with velocity resolution $\sim$4 \kms\ and
spatial resolution $\sim$1.2\arcsec\ along a 1.4$\arcsec\times$7.5\arcsec\ slit. 
The data were first processed with the standard TEXES pipeline reduction program.
The absolute astrometry of the 12.8$\mu$m data is a  few arcseconds. We obtained coordinates for the [Ne
II] maps by aligning them with the radio continuum  maps with positional errors of  $\sim0.35''$ (1 pixel).  
This is because the position of the strong radio source Sgr A* is accurately known in radio continuum images
taken with the VLA.  

\subsubsection{Mid-IR data}

We used the published data obtained with  the Very Large Telescope (VLT) at wavelengths ranging between 1.6 and 19.5 $\mu$m 
(Viehmann et al. 2006). 
This data set comprised of  high resolution and high sensitivity observations 
of stellar sources  near Sgr A*. 
The data points for Spectral Energy Distribution (SED)
fitting (see $\S3.2$) are taken from Table 1 of Viehmann et al. (2006)\footnote{The 
relative coordinates of stars, which are misprinted in their Table 1, are with respect to IRS~7 at 
$\alpha, \delta \; (J2000) 17^h 45^m 40.040^s, -29^{\circ} 00' 22.56''$ (private communication)}.

\section{Results and Discussion}
\subsection{Water Masers} 

We detect 13 water masers concentrated toward the inner edge of the molecular ring. Figure 1 shows the 
distribution of water masers (crosses) superimposed on an image of the velocity-integrated HCN (1-0) line (Christopher et 
al. 2005); 
The insets 
show their spectra. Entries in columns 1 to 7 in Table 1 give Galactic coordinates, source name, celestial 
coordinates, the peak flux, the peak velocity and cross references for  all detected masers. 
A large-scale water maser survey 
toward the Galactic center using the Australia Telescope Compact Array (ATCA) did not detect any of the sources reported 
here because of their lower  sensitivity with rms noise $\sigma\sim0.1$ Jy (Caswell et al. 2011). Table 1 lists multiple 
velocity components
 in several sources with peak velocities ranging between $\sim-91$ and 67 \kms. The peak fluxes range  between 22.6 and 
319.6 mJy beam$^{-1}$\,  corresponding to brightness temperatures ranging  between $\sim0.6$ and 7.9K. 
These sources are spatially 
unresolved, thus these brightness temperatures are lower limits.  We regard these sources as masers because of 
their narrow linewidths of $\sim$1 \kms.

The HCN (1-0) observations resolved 26 cores with a typical diameter of  $\sim7''$ (0.25 pc). 
Three collisionally--excited methanol masers  coincide with 
three of 26 HCN  clumps F, G and V, drawn as circles centered on masers in Figure  1 (YBWR08). 
These clumps are  
The region to the NE of the molecular ring in Figure 1 shows three water masers (1, 11 and 12) 
and two methanol masers coincident 
with the clumps labeled F and G,  in the HCN (1-0) line map (Christopher et al. 2005; YBWR08).
 The methanol spectrum in the vicinity of clump F 
has  three narrow velocity peaks between 52 and 55 \kms\, and a broad red-shifted wing extending up to 100 \kms 
(YBWR08). This region is inferred  to be a star forming site because of the 
collisionally excited methanol 
maser at 44 GHz and the  red-shifted broad wing in the HCN (1-0) line with a peak velocity 
of $\sim60$ \kms\, (YBWR08).  The velocity correlation of 22 GHz masers and molecular line  emission, SiO (5-4) and HCN (1-0),  
will be discussed in $\S3.3.3$.

It is possible that interstellar water lines are contaminated by water lines from evolved OH/IR stars in this highly confused region 
of the Galaxy. To exclude this possibility, we compared the position of new water masers listed in Table 1 with those of OH/IR stars 
and SiO masers associated with evolved stars. With the exception of sources 1 and 13 (Lindqvist et al. 1992; Levine et al. 1995; 
Sjouwerman \& van Langevelde 1996; Sjouwerman et al. 2002; Li et al. 2010), we did not find evolved stellar counterparts with similar 
velocities and positions to those of detected water masers in Table 1. Water maser 1 coincides with an OH/IR star 359.956-0.050 2a 
(Table 2 of Sjouwerman et al. 2002), although it is not clear why this water maser has 
four velocity components 47, 54, 58, 63 \kms\, 
(Table 1) when the velocity profile of the OH/IR star has only two velocity components (Sjouwerman et al. 2002). These velocity 
components match up with those of collisionally excited methanol line emission (YBWR08). Similarly, maser 13 also shows five 
velocity 
peaks 78.53, 80.22, 81.90, 87.81, 91.19 \kms\, (Table 1) 
and is spatially coincident with the OH/IR star 359.970-0.049, with two velocity peaks 37.5 and 94.5 \kms\, (Sjouwerman et al. 
2002). The spectrum of maser 8 appears to be similar to that of an OH/IR star.

Among other water masers, we note that water maser 11 shows two forbidden velocities $-63$ and $-71$ \kms\, assuming orbits aligned with 
the Galactic rotation. An SiO maser source,  denoted SiO-18 in Table 2 of Li et al. (2010) is located within a few arcseconds of
water maser source 4. However, the peak emission from SiO-18 is at 39.6 \kms\, whereas water maser source 4 has a peak velocity of $-82$ 
\kms\, suggesting that they are not associated with each other. We note water maser 6 having more than three peaks like 
water maser 1 and 
13.  OH/IR stars are identified by two peaks symmetrically centered around the velocity of the central star. 
So, water masers with more than two velocity components are unlikely to be associated with 
individual OH/IR stars. 
It is possible that these sources with multiple velocity components are interstellar water masers but 
are  contaminated by OH/IR stars. 
Water maser 8 has two components that are much more widely 
separated in velocity than other water masers' components.
In summary, we have detected  a total of 5 water masers 
with three or more   (1, 5, 6, 11, 13)  velocity components, five water masers  with single velocity 
component (3, 4, 9, 10 and 12) 
and  three water masers with two velocity components (3, 7 and 8).  
 Water masers with double velocity components are probably  associated with 
OH/IR stars whereas the remaining sources are likely to be associated with 
interstellar  masers.

\subsection{YSO Candidates and Star Formation Rate} 

The detection of water and collisionally excited methanol masers motivated us to search for YSOs within the inner pc of Sgr 
A*. This region is populated by  about 60   dusty and red sources which could be OH/IR stars and/or YSO candidates, both of 
which have strong infrared excess emission (Viehmann et al. 2005, 2006).  OH/IR stars represent evolved AGB stars at the end 
of their lives (Habing  1996) whereas YSOs are characterized by dusty envelopes and disks that absorb the radiation 
from the central protostar (e.g., Whitney et al.  2003). Near and mid-IR photometric observations with the 
VLT  have 
identified a total of 64 sources in the central parsec at wavelengths 
ranging between 
1.6 $\mu$m (H band) and 19.5 $\mu$m (Q band), 
respectively (Viehmann et al. 2006). 
Four types  of sources were classified based on their SEDs
 (Viehmann et al. 2006; Perger et al. 2008). 
Type I consists of luminous IR sources that are embedded in the 
N arm ionized streamers. These sources (e.g., IRS~1, IRS~2L, IRS~10W and IRS~21) 
showing bow-shock structures with featureless 
SEDs that peak in the mid-IR wavelengths,  are generally thought to be interacting with the mini-spiral (Tanner et 
al.  2005). 
Type II consists of low luminosity sources that have similar 
 bow shock morphology to those of Type I sources, but 
are offset from the mini-spiral streamers. 
Type III sources are cool stars with SEDs 
that peak in the near-IR wavelengths. Lastly, Type IV sources are hot stars (e.g., IRS~16NE, IRS~16NW, AF and AHH) with 
SEDs that peak at shorter wavelengths. There are also a number of sources that could not be classified among these four 
types based on their SEDs (Viehmann et al. 2006).

%One of the difficulties  of  identifying a dusty source by their colors is that
%the SEDs of hot dusty and bright AGB stars can not be easily identified. 
%we fitted SEDs of dusty stars  and identified  
%sources that are consistent with YSOs tracing signs of star formation on time scales of 
%$\sim10^{5}$ to few times 10$^5$ years.  

We carried out  YSO SED fitting of dusty sources of all types 
listed in Table 1 of Viehmann et al. (2006) to provide additional 
constraints on the nature of these sources. We used the same technique that identified a population of YSO candidates 
toward the inner few hundred parsecs of the Galaxy (Yusef-Zadeh et al. 2009, but  see 
Koepferl et al. 2014). 
To obtain reliable fits, we only selected 
sources with at least four flux measurements (52 out of 64 sources). The Viehmann et al. (2006) observations used nine 
VLT/VISIR bands in the wavelength range from 1.6 $\mu$m to 19.5 $\mu$m, with an angular resolution (FWHM) of 0.3$''$ -- 
0.6$''$.  We considered using the catalog of near-IR 
sources (Sch\"odel et al. 2010) for 
additional data points for those sources that had limited data in Viehman et al. (2006). 
However, the lack of proper motion 
data and confusing sources 
prevented us from using the  new catalog at H, J and L$'$ bands (Sch\"odel et al. 2010). 
The availability  of  data from a single instrument 
that  provides a broad wavelength range and high angular 
resolution minimizes source confusion and systematic effects.

We compare the sources' SEDs to the Robitaille et al. (2006) large grid of pre-computed YSO model SEDs using the linear 
regression SED fitting tool developed by Robitaille et al. (2007).  We consider models with 
interstellar extinction ($A_{V}$) between  25 and  30 mag, and distance between 7.65 and  9.35 kpc (i.e., 
the distance to the Galactic center of 8.5 kpc $\pm$10\%).  We  identify 
 well-fit 
models as those with a  normalized $\chi^2$ per data point ($\chi^2/pt$) less than  3.  As a result, 
only 19 out of 52 sources are  well-fit by  YSO models.
A good YSO model fit indicates that the source's SED is consistent with the YSO model SED, 
but  does not provide a definite  classification.

%({\it it may change, I am still trying to figure out the difference between the old and new fitter}) 

We estimate physical parameters of these sources by averaging parameters of all YSO models that fit the source's SED 
with $\chi^2/pt$ in a range between the $\chi^2/pt$ for the best-fitting model ($\chi^2_{min}/pt$) and 
($\chi^2_{min}/pt$ + 2). The visual examination of the fits showed that the YSO model fits with $\chi^2/pt$ within this 
arbitrary range  provide acceptable fits to the SEDs (see Figure 2). 
The fitting results are presented in Table 2. In Column 1, we provide
the Viehmann et al. (2006) ID numbers (V06 ID). Columns 2 and 3 list the
$\chi^2/pt$ for the best fit model ($\chi_{\rm min}^{2}/$) and the
number of acceptable fits ($n_{fits}$), respectively. In Columns 4--15,
we provide the estimates of the physical parameters for the best fit
model, as well as the average physical parameters calculated based on
all the acceptable fits (see above). The physical parameters include the
interstellar extinction ($A_{V}$), stellar luminosity (L$_{\star}$),
temperature (T$_{\star}$), and mass (M$_{\star}$), as well as the
envelope mass (M$_{env}$), and disk mass (M$_{disk}$). The best and
average evolutionary stages (Columns 16--17) are based on sources'
physical parameters as defined by Robitaille et al. (2006).  Column 18
gives the number of data points used for the SED fitting.

%Table 2 lists ID numbers in column 1, taken from Viehmann et al. (2006), and the $\chi^2_{min}/pt$ of the best fit model and 
%the number of acceptable fits in columns 2 and 3, respectively. We note that source 5 in Viehmann et al. (2006) is equally 
%well-fitted by stellar atmosphere model. Estimates for physical parameters, namely, the interstellar extinction (A$_v$), 
%stellar luminosity in units of 10$^4$\lsol, stellar temperature in units of 10$^4$K, stellar mass, envelope mass and 
%evolutionary stages derived based on sources' physical quantities, as defined by Robitaille et al. (2006), are given in 
%columns 4 to 17, respectively. The best fit value and the average value with one $\sigma$ rms error for each physical 
%quantity are given in two separate columns.  Column 18 gives the number of data points used for SED fitting.

Stage I objects are young protostars embedded in an opaque infalling envelope. In Stage II objects the envelope has 
mostly dispersed, and the central star is surrounded by an opaque disk. In Stage III the disk is optically thin.  The 
SEDs of sources that are well-fit with YSO models are shown in Figure 2. In each plot, the best fit model and all 
acceptable models are shown in black and gray, respectively. The dashed line is the central stellar atmosphere 
corresponding to the best fit model, extincted by the fitted foreground extinction. To place additional constraints on 
best fit models, we explored the possibility of using Herschel data at long wavelengths. However, the Herschel PACS 
instrument (Poglitsch \etal\, 2010) had insufficient angular resolution ($\sim5.5"$) at 70$\mu$m to resolve individual 
YSO candidates in this crowded field and in the presence of diffuse emission from the CMR. Despite this, for some 
sources, the generous upper limits that are derived, ranging between (3.24--5.75)$\times10^{-9}$ erg\, s$^{-1}$ 
cm$^{-2}$ for sources listed in Table 2, can rule out a few of the models with the most massive envelopes. Somewhat 
more restrictive upper limits (e.g. factors of ~10 lower) would eliminate models with envelopes greater than $\sim14$ 
\msol\.  Sources 10, 11, 26, and 39 would be most strongly impacted, changing them from likely Stage I objects to 
Stage II objects.

%The uncertainties of the average physical parameters represent standard deviations of the mean. 

SED modeling of the infrared excess sources listed in Table 2 indicates that they are massive YSO candidates. 
In particular, the SED fits to members of the two complexes, IRS~5 and IRS~13N, 
are consistent with being massive YSO candidates.
Four members of the IRS 5 Complex (IRS 5, IRS 5E, IRS 5S and IRS 5SE1), 
as well as sources IRS 13N, IRS 13NE, IRS 2L, IRS 1W, and IRS 10W are
classified as main sequence or WR stars that may be associated with
ionized streamers. 
Stellar sources  in the IRS~5 Complex 
have  been interpreted as  low-luminosity bow shock sources, low-luminosity dust 
forming AGB stars or YSOs (Preger et al. 2008). 
Based on proper motion measurements of cluster members in IRS~5 and IRS~13 as well 
as their high luminosities, these sources are unlikely to be AGBs and could be members of young cluster of 
stars (Preger et al. 2008; Eckart et al. 2013). 
The distribution of luminous YSO candidates imply that  on-going star 
formation is taking place near Sgr A* within the  
ionized streamers orbiting Sgr A* (e.g., 
Viehmann \etal 2006; YNWR08; Eckart et al.  2013; Nishiyama and Sch\"odel 2013; Yusef-Zadeh et al. 2013, 2014; Jalali 
et al. 2014).  
Additional support for ongoing star formation within 2pc of Sgr A* comes from  radio continuum and 
near-IR polarization studies, as described below.  
Radio continuum observations did not detect a bow shock structure but 
resolved IRS~21 into four components (Yusef-Zadeh et al. 2014). 
The radio emission from a cluster of four YSOs is 
due to thermal bremsstrahlung from ionized gas that is being           
photo--evaporated from a disk by the strong UV radiation field at the Galactic center

In our sample of YSO candidates we include two sources that were not well-fitted with the YSO models. These are 
IRS 1W and IRS 10W with $\chi_{\rm min}^{2}/$ of 12 and 5, respectively. Their SEDs are shown separately in
Figure 3. These sources are not included in Table 2 since the physical parameters estimated from the fitting are
highly uncertain. However, other observations indicate that these sources are likely YSO candidates.  IRS 1W and
IRS 10W coincide within three clumps of   SiO (5-4) emission 1,2 and 3 (Yusef-Zadeh et al. 2013). 
SiO (5-4) line emission with broad linewidths, as discussed in 
$\S3$, providing  strong support for protostellar outflows interacting with  the ISM.  
Additional 
support comes from near-IR polarization measurements indicating that sources 
IRS~1W, IRS~10W, IRS~21 are intrinsically polarized  
sources  (Yoshikawa et al. 2013). 
Intrinsically polarized sources are 
argued to be associated with YSOs driven by outflows in  star forming regions (Yoshikawa et al. 2013).
Table 3 gives the Viehmann et al. (2006) ID numbers (V06 ID), the corresponding near-IR names (Names), and 
equatorial coordinates for 19 well-fitted YSO candidates, as well as sources IRS 1W and IRS 10W. Cross-references
are provided for some of the sources.

Recent radio continuum observations within 30$''$ of Sgr A* show evidence of partially resolved free-free emission
from near-IR identified stellar sources, e.g., IRS~21, IRS~5, IRS~13N and IRS~13E (Yusef-Zadeh et al. 2014). We
argued that the radio emission is due to thermal bremsstrahlung from ionized gas that is being photo--evaporated
from a disk by the strong UV radiation field at the Galactic center. In this picture, the mass-loss rate due to
photo--evaporation implies the existence of a reservoir of neutral material. Robitalle et al (2006) models consider
envelope outer radii ranging between $\sim1.5\times10^{16}$ to $\sim1.5\times10^{18}$ cm. This range in size
appears to be motivated by the limit at which the temperature would drop to 30K. On the other hand, disk and inner
envelope radii from our SED models range between 5 and 3000 AU (7.5$\times10^{13}$ and 4.5$\time10^{16}$ cm). In
the case of evolved stars in the Galactic center region, there is no identified AGB star that is detected at radio
wavelengths. The outer radii of AGB stars can be as large as few $\times10^{17}$ cm (Wallerstein \& Knapp 1998). If
the envelope of AGBs or YSOs were photoionized, their angular size at the Galactic center would show extended radio
sources greater than 1$''$ (1" is 0.04 pc at the distance of the Galactic center). One of the dustiest AGB sources
in the Galactic center is IRS~3 (Preger et al. 2008) which shows no signature of extended thermal emission. Radio
sources are partially resolved at $\sim0.1''$ resolution. Thus, detection of partially resolved radio continuum
sources in the Galactic center is consistent with SED models of infrared excess sources. Notable sources indicating
a disk of material associated with YSO candidates are IRS~1W, IRS~ 5, IRS~5NE, IRS~13N, IRS~13NE, IRS~21 and
IRS~34SW corresponding to sources 1, 13, 14, 47, 48, 54 and 57 in Table 2.

Some of the  stars, IRS~1W, IRS~10W, IRS~5, IRS~2L and IRS~21, show bow shock structures with featureless SEDs that peak in 
mid-IR wavelengths (Clenet et al. 2004; Tanner et al. 2002, 2005; Muzic et al. 2010; Sanchez-Bermudez et al. 2014). The bow 
shock morphologies  of these sources are interpreted in terms of 
 the interaction of ionized winds with the streamers of ionized gas 
(Tanner et al. 2005; Sanchez-Bermudez et al. 2014). This model adopts a spherical outflow from central Wolf-Rayet stars 
(Tanner \etal 2002, 2005). 
 Alternatively, the interaction of ionized streamers could take place with the material evaporating 
from the YSO disks assuming that 
the outflow from the YSO disks is faster than the relative 
motion between the YSOs and the ionized streamers. 
In this scenario, 
the cause of the asymmetry comes from the outflow arising  preferentially from the disk surface, facing the ionizing 
sources  in the central cluster of stars. 
The asymmetry in the bow shock geometry could also  come from YSO
disks interacting with ionized streamers. 
The orientation of  YSO disks should be random in the disk picture. 
The orientations of the bow shock 
in IRS~10W, IRS~5 do not seem to follow the north-south direction of the streamers,  thus 
consistent with  disks associated with  YSO candidates. 

YSO candidates' stellar masses derived from the SED fitting can be used to estimate the star formation rate (SFR; 
e.g., Povich et al. 2011; Whitney et al. 2008; Yusef-Zadeh et al. 2009). The SFR is calculated by estimating the total 
mass of the YSO candidates and dividing it by YSOs' approximate lifetime. Since our sample of YSO candidates is small, 
we can only provide a rough approximation of the SFR. We use Stage I (the average evolutionary stage from Table 2) YSO 
candidates only to get a sample of sources at approximately the same age and lifetime.

To estimate the total mass of the YSOs, we construct the stellar mass function, scale the Kroupa (2001) initial mass 
function (IMF; $dN/dm \propto m^{-\alpha}$, m -- a stellar mass) to match the peak of the distribution, and integrate 
under the IMF over a mass range of 0.08--50 M$_{\odot}$. The Kroupa (2001) IMF has a slope ($\alpha$) of 1.3 between 
0.08 and 0.5 M$_{\odot}$ and 2.3 between 0.5 M$_{\odot}$ and 50 M$_{\odot}$. The estimated total mass of the YSO 
candidates is $\sim$1318 M$_{\odot}$. Assuming a typical lifetime of the Stage I YSO of 10$^{5}$ years, the SFR within 
1 pc of Sgr A* is estimated to be $\sim$1$\times10^{-2}$ M$_{\odot}$ yr$^{-1}$. 
This is only an order magnitude estimate because of the small 
number statistics, uncertainty of a factor of 2 in the age of Stage I YSOs (see Koepferl et al. 2014), and the 
extrapolation of the IMF down to the cut-off at 0.08 M$_{\odot}$

\subsubsection{YSO candidates in Rotating Disks Orbiting Sgr A*} 

One of the most intriguing  aspects of the YSO candidates  is their spatial 
distribution which appears to be   similar to that  of the young, hot  stars orbiting Sgr A*. 
The so-called 
edge-on clockwise and face-on counterclockwise stellar disks  have  positive and negative 
angular momentum j, respectively 
(Beloborodov et al. 2006; 
Paumard et al. 2006) and consist of 
hot and massive stars  which are  few million years old (Paumard et al.   2006; Lu et al. 2009). 
The two disks extend out to  10$''$ from Sgr A*. 
The position angles (PAs) of young stars  
in  the clockwise and counterclockwise stellar disks 
are $\sim 30^{\circ}$  and $130^{\circ}$, respectively.  
Figure 4a,b show the positions of YSO candidates superimposed on a 34  GHz continuum  and 3.8$\mu$m               
images, respectively. YSO candidates IRS~5, source 28 and IRS~2L are embedded or               
offset to the east of the N arm of the ionized minispiral, 
IRS~4 and sources 12 lie in the E arm and IRS~13E, IRS~13N and IRS~34SW 
lie in projection against the  bar of ionized gas and its NW extension.
The YSO candidates that are  distributed  diagonally from NE to SW run parallel to the edge-on clockwise  
disk whereas the sources to the SE fall along the counterclockwise stellar disk. 
The distributions of YSOs extend to $\sim15''$ from Sgr A*,  larger than the 10$''$ extent of 
counter rotating disks. 
The three dimensional motions of many of the YSO candidates 
are unknown, however,  the line of sight motion of the  YSO candidates  and their apparent  distribution on the sky 
show a trend that is consistent with the positive and negative 
angular  momenta of two rotating disks. 
For example, the proper motion of
the  IRS~5 cluster suggests that the overall angular  momenta of these sources is 
in agreement with the clockwise rotating disk of stars (Preger et al.
2006).  In addition, IRS~34SW has positive angular momentum, thus follow 
clockwise rotating disks whereas IRS~13 has negative angular momentum values. 
This suggests that  the YSO candidates are forming in the residual 
outer regions of the original gas disks that formed the original stellar disks orbiting Sgr A*.
Future proper motion measurements  will determine the  angular momentum of the YSO  candidates 
and test the picture proposed here. 

%Where known, the  angular momentum (j) of  YSO candidates are  shown in column 10 of Table 2. 

\subsection{Protostellar Outflows Traced by SiO Emission} 

Another indication that the region within  0.5 pc of Sgr A* is a site of ongoing star formation comes from 
clumps of SiO emission  found in this region (Yusef-Zadeh et al. 2013). 
ALMA observations of the interior of the molecular ring detected 11 clumps of  SiO (5-4) 
emission  within the inner 0.5 pc of Sgr A*. 
SiO line emission  is generally associated with  shocks from protostellar 
outflows, so  SiO clumps detected within the molecular ring were interpreted as 
evidence of 
young star formation activity  near Sgr A* where outflows from YSOs interact with the surrounds. 
The location of five of SiO  clumps  are  offset from  the
eastern edge of the N arm but run parallel along the N arm of ionized gas. 
These sources  show typical linewidths  between 11 and 21 \kms and peak
radial velocities that decrease from 37.8 \kms\, to 7.4 \kms\, to the N of the streamers. 
This trend in
radial velocity  is consistent with the trend  in the kinematics
of ionized gas of the N arm (e.g., Zhao \etal 2010). However, the radial velocity distribution of SiO clumps does  not match 
the kinematics of ionized gas in the N arm. 
The positions  of two of the most prominent SiO (5-4) clumps found by ALMA (sources 1 and 11 in Yusef-Zadeh et al. 2013)
lie  along the  extension  of YSO candidates found to the NE where the IRS~5 cluster is located. 
This suggests that  there is a concentration of molecular gas to the east of the N arm.

We  extended the study of SiO (5-4) line emission from the inner 0.5 pc to the molecular ring within
$\sim$2 pc of  Sgr A*  using data taken with the SMA with 
a resolution of $3.6''\times2.4''$
(Martin et al. 2012). Table 4 lists the parameters of Gaussian fitted SiO (5-4) line emission  found in 
the molecular ring. 
Columns (1) to (7) show  the source number,  following the SiO (5-4) sources detected by ALMA (Yusef-Zadeh et al. 2013), 
celestial coordinates, peak intensity, FWHM and 
full 
width at zero intensity (FWZI), respectively. 
The SiO (5-4) sources  detected by  ALMA are concentrated in a  region within  the molecular ring  where 
SMA observations could not detect any SiO (5-4) line because of its poor  sensitivity 
with the exception of 
source 10 which  was detected in both ALMA and SMA observations. 
Table 4 shows the parameters of the fit to this source detected with ALMA and SMA  
are similar to each other. 
The discrepancy between the fitted parameters of 
SMA source 10 and  10 (ALMA) in Table 4 is mainly due to different spatial and spectral resolutions and sensitivity of 
the ALMA and SMA data. 

Figure 5 shows 
the velocity integrated SiO (5-4) line intensity of  26 clumps  toward  the molecular ring.  
To investigate if these  new SiO clumps 
have similar  characteristics to known protostellar outflows in the Galactic disk, 
Figure 6a compares the linewidths and SiO (5-4) luminosity of the emission from 11 clumps interior to 
the ring (red), found from 
ALMA observations, 26 clumps (12 to 36) in the molecular ring listed in Table 4 (blue) 
as well as low and high-mass protostars in the Galactic disk (black). 
The SiO   sources lying in the central hole in the central molecular ring  as well as the clumps 
coincident with the ring show 
a similar lack of  dependence between the FWZI and SiO luminosity  found for 
high mass protostellar systems in the Galactic 
disk (Yusef-Zadeh et al. 2013).  This supports the suggestion that the SiO clumps in the 
CMR  are  massive protostellar outflows  and the CMR itself is  a site of recent massive star formation. 
We note a bipolar outflow candidate (Fig. 6b) in one of the SiO clumps, as described below. 

\subsubsection{SiO (5-4) Bipolar Outflow} 

One of the clearest signatures of ongoing star formation activity is 
bipolar outflows traced by  
SiO line emission. We identified  a bipolar outflow  candidate 
in  clump F or clumps 22 and 23 in Figure 5 of the CMR. The reason that we were able to identify this source is because the F clump 
is less confused kinematically and spatially than the material in the molecular ring. 
Figure 6b shows contours of SiO (5-4) line intensity for a velocity range  between 30--55 \kms\, (blue) and 65--90 \kms\, (red). 
The kinematics of SiO (5-4) line emission  show the appearance
of a bipolar outflow from clumps 22 and 23. The spectrum of this source shows broad  blue and red-shifted wings and 
the blue and red-shifted radial velocities along the linear features are reminiscent of bipolar outflows. 
The direction of the velocity gradient of bipolar lobes is the 
opposite to that seen in the molecular ring (Christopher et al. 2005).  
The presence of methanol and water masers and bipolar outflow in the vicinity of clump F 
provides  the strongest  evidence for   ongoing star formation in the molecular ring. 

\subsubsection{LVG Analysis of SiO Lines} 

In order to apply an LVG analysis to the sources detected in the SiO (5-4) and (2-1) lines, we first convolved the 
CARMA and SMA data to an identical resolution of $6.85''\times3.73''$ and then performed Gaussian fits to the sources, 
obtaining the parameters listed in Table 5. The CARMA sensitivity of SiO (2-1) was not sufficient to detect source 12. 
In addition, SiO (5-4) and SiO (2-1) data from SMA and CARMA showed some clumps with multiple velocity components. The 
LVG analysis adopted a uniform slab geometry for the emitting region, and included a radiation field characterized by 
a dust temperature of 75 K, 30 magnitudes of visual extinction and a dust extinction curve following Draine \& Lee 
(1984). We computed the emergent intensities in the two lines given the slab's assumed gas density, temperature, full 
width at half maximum (FWHM) velocity, and SiO column.  In Figure 7, we show contours of 5-4 vs 2-1 brightness 
temperature for constant hydrogen number density ($n_\mathrm{H}$) and for fixed SiO column density, assuming a FWHM of 
30\,km\,s$^{-1}$ for kinetic temperatures of 150\,K. Also plotted are the points corresponding to the observed SiO 
sources which lie close to and follow contours within $n_\mathrm{H} = 10^6 $ or n(H2)$\sim 5\times10^5$ and $10^{6.15} 
$\,cm$^{-3}$ for the assumed kinetic temperature of 150 K. These estimates are similar to those given by Amo-Baladron 
et al. (2009) Note that the predicted brightness temperatures scale linearly with SiO column density because the lines 
are optically thin.  We conclude that the sources have $n_\mathrm{H} \approx 1-4\times10^6 $\,cm$^{-3}$ for assumed 
kinetic temperatures of 50 and 150 K and $N(\rm SiO)\approx 10^{14.4}\,\mathrm{cm}^{-2}$.  As with other linear 
molecules, we find that the density inferred from the LVG analysis is inversely proportional to the assumed kinetic 
temperature, so that the LVG constrains the pressure in the SiO sources to be $P/k \approx 1\times 10^{8}\, 
\mathrm{K\,cm}^{-3}$.

%NEED TO CHECK the NUMBERS

The SiO clumps are interpreted to be arising from protostellar outflow sources. Using a typical molecular density 
$\sim10^6$ cm$^{-3}$ and a radius of $\sim2''$, the total mass that is swept up a molecular clump of SiO emission by 
protostellar outflow is estimated to be $\sim$68 \msol.  The kinetic luminosity L$_{kinetic}=0.5\, \rm {M}\, \rm v^2$ 
of the swept up material over the dynamical age $\sim$1300 years is also calculated to be $\sim$500 \lsol. Using the 
estimated SiO and molecular hydrogen column densities, the abundance of SiO in each SiO clump is estimated to be 
2.5$\times10^{-10}$. The column density of hydrogen which itself is estimated by multiplying the gas density and the 
size of a clump.  However, SiO shocks are generally associated with a thin layer where the protostellar jet interacts 
with a cloud.  So, we overestimated the thickness of the SiO layer by a factor of 10 to 100 and therefore 
underestimated the SiO abundncae.  Thus, the SiO abundance could range  between  2.5$\times10^{-9}$  and 
2.5$\times10^{-8}$.

%We have now clarified that our estimate is a lower limit and could easily
%increase by a factor of 10.

\subsubsection{Water Maser, HCN and  SiO  Lines} 

To investigate whether  water masers (see Fig. 1) are tracking star formation activity, 
we compare the distribution of SiO (5-4) emission with the spectra of several water masers toward the molecular ring.   
Figure 8 shows 
a grayscale image of SiO (5-4) line emission from the molecular ring and the insets show six SiO (5-4) spectra from 
regions where water masers have been detected. Water masers 1, 11 and 12 lie in the vicinity of HCN(1-0) clumps F and 
G, whereas water masers 4 and 5 are concentrated near HCN (1-0) clump O. These SiO features associated with clumps F, G and 
O to the E and SW of the molecular ring are resolved spatially in the NS direction. 
The  SiO (5-4) spectra 21 to 24 from the 
HCN (1-0)  clumps F and  G  reveal velocities peaking between 30 and 60 \kms. 
These velocities  are generally consistent with those of water masers 1 and 12. 
The  velocity profile of SiO (5-4) from clump V shows two blue and red-shifted velocity  components
whereas HCN (1-0) spectra shows a broad asymmetric red-shifted wing.  
SiO (5-4) line emission from Clump O 
shows  a velocity peak at $\sim-8$  \kms\,  and extends to negative velocity of --100 \kms.  The broad linewidth of 
this clump includes the LSR velocities of water masers 4 and 5. 
The spectra of SiO (5-4) line emission from  clump O  exhibits an asymmetric blue-shifted wing. 

Previous H$_2$ observations of the molecular ring (Gatley et al. 1986; Yusef-Zadeh et al. 2001) have identified 
shocked H$_2$ features associated with clumps V and O.  The correspondence between SiO, HCN(1-0) and H$_2$ molecular line 
emission suggests  that H$_2$ emission from the CMR is shock excited. 
The velocity profiles of SiO emission 
are similar to those of HCN thus suggesting that SiO (5-4) and HCN (1-0) line emission arise from the same region. In 
addition, velocities of water masers 1, 12, 4  and 5 are within the range of SiO emission from clumps F, G and  O, thus 
suggesting that these water masers may be associated with the molecular gas in the ring. 

%The SiO (2-1) line intensities of 
%clumps F, G, V and O integrated over all velocities are 25.5, 21.0, 1.7 and 72.3 Jy\, \kms, respectively.

%The integrated SiO (2-1) line emission from the entire region is estimated to be $\sim$253 Jy\, \kms.

% STAR FORMATION ATIVITY FROM THE INTEGRATED SIO FLUX
%which we call clump JJ, 

%There is no evidence of  H$_2$ emission  from clumps F and G. 

\subsection{{[Ne II] Line Emission}} 

Figure 9a shows contours of HCN (1-0) line emission at the peak velocity of clump V superimposed on a grayscale 
continuum image of the mini-spiral HII feature at 5 GHz (Yusef-Zadeh \& Wardle 1993). 
A radio continuum ionized feature, which we refer to as the EW Ridge, extends away from clump V at roughly constant 
declination. 
This  ridge of ionized gas branches into  two linear features, each with 
horizontal and vertical extent of $\sim20''$ (0.8 pc) and 1$''$ (0.04 pc), respectively, crossing each other with position 
angles of $81^{\circ}$ and 105$^{\circ}$ at the peak of clump V peaks. The kinematics of this feature do not follow the 
kinematics of the molecular ring. Morphologically, the linear features wiggle to the east as they merge with the N arm of 
the mini-spiral.  Figure 9b shows the 
position velocity (P-V) contour plot of Ne II line emission along the brighter branch along 
the length of the EW 
ionized ridge at a position angle of $-93^{\circ}$ at almost constant declination. 
This plot illustrates the gradual change in the velocity of the linear 
feature from $\sim55$ to --55 \kms\, with a velocity gradient of $\sim$470 \kms\, pc$^{-1}$. The red and blue-shifted 
velocity components are associated with the eastern and western halves of the crossed linear features, respectively. Given 
the high velocity gradient of Ne II line emission, the ionized gas associated with the EW  ridge is kinematically disturbed 
and is likely to be unbound
to the Galactic center gravitational potential. 
The blue and red-shifted radial 
velocities along the linear features are reminiscent of bipolar outflows.  

%CHECK THE NUMBERS BELOW

We suggest that the EW feature is a result of jet activity from two protostars in clump V. A collisionally excited 
methanol maser is associated with clump V (YBRW08), at a velocity of $\sim40$ \kms, superimposed on broad thermal line 
emission extending to velocities of $\sim80$ \kms. There is no evidence of water masers arising from clump V.  The 
dynamical age of the outflow is estimated to be $\sim1\times10^4$ years using the size of the ionized ridge $\sim 20''$ 
(0.78 pc) and the difference in the blue and red-shifted velocities $\sim80$ \kms. The intensity of the emission from 
the EW Ridge is $\sim$0.5 mJy beam$^{-1}$ at 22.4 GHz with a resolution of $0.24''\times0.17''$.  The background 
subtracted flux density of the ionized EW Ridge is estimated to be $\sim0.5\pm0.2$ Jy at 22.4 GHz. The ionized gas is 
estimated to have an emission measure EM$\sim6\times10^6$ pc cm$^{-6}$ corresponding to an ionized gas density of 
8.7$\times10^3/f^{0.5}$ cm$^{-3}$ where $f$ is the volume filling factor. 
The electron  temperature is assumed to be T$_e\sim8000$K and a path-length L=0.08 pc which is 
equivalent  to the jet diameter $\sim2''$. The total mass of 
the EW Ridge and the mass loss rate of the outflowing ionized material are estimated to be 
$\sim1.7\times f^{0.5}$ \msol\, 
and  $\sim1.8\times f^{0.5}\times10^{-4}$ \msol\, yr$^{-1}$, respectively. 
Assuming 
10\% lumpiness in the density of ionized gas, $f=0.1$, the 
total mass and the mass loss rate are
$\sim0.5$ \msol\, and  $\sim5\times10^{-5}$ \msol\, yr$^{-1}$, respectively.  These values are generally consistent 
with those found in  star forming regions.

\subsection{Radio Dark Clouds within 1 pc  of Sgr A* } 

Star formation activity within the inner 2 pc of Sgr A* require 
a reservoir of molecular gas surrounding  the sites of young star forming activity. 
The CMR provides the largest concentration of molecular gas near Sgr A*. 
Recent observations have also shown additional molecular  material 
is distributed within the CMR, 
contrary to the widely accepted idea that a cavity of mini-spiral shaped ionized gas 
fills inside the molecular ring (Jackson et al. 
1993; Monterrey-Cast\~no et al. 2010; Martin et al. 2012;
Yusef-Zadeh et al. 2013).
 Here we provide a new tracer of  extended  neutral gas by identifying radio dark clouds 
within the molecular ring. The presence of radio dark clouds is consistent with 
a reservoir of neutral gas that feeds  star formation activity close to Sgr A*. 

%In addition,   high resolution radio continuum images
%show  two layers of photoionized  gas sourrounding  cold gas 
%residing  within the molecular ring. 

Recent radio images of the Galactic center have revealed a large number of radio dark clouds (Yusef-Zadeh 2012). These 
features coincide with cool neutral clouds embedded in a hot ionized medium, tracing volumes  where there is a 
deficiency of free-free radio continuum emission.
Figures 10a,b  show the distribution of ionized gas surrounding Sgr A* at 7mm (Yusef-Zadeh et al. 2014). 
We note  three dark features, two near the N and E arms of the mini-spiral and one near the IRS~13 complex. 
The darkest features  lie  to  the east of IRS~13E and SE of IRS~13N, identified by a red circle with a diameter $\sim1''$. 
We suggest  that these dark features are associated with radio dark clouds. Interferometric errors 
due to incomplete sampling of {\it {uv}} plane also produce dark features. 
However, the regions labeled in Figure 10 have been detected in radio images at multiple wavelengths 
based on different observations, suggesting that they are real features. 
Future high-resolution molecular line observations should confirm our interpretation of radio dark clouds.  
The mean flux density averaged over the red circle near IRS~13 is 
fainter by 0.1-0.2 mJy than the region immediately outside the circled region.
 The difference
between the flux density of the ambient gas toward and away from the cloud suggests that the depth of the ionized layer 
 is 5-10 times bigger than the dimension of the radio dark cloud.  This radio dark cloud may be tracing the molecular material
associated with ongoing star formation near IRS~13N. Similarly, the radio dark cloud near the N and E arms indicate that the
ionized gas in the mini-spiral feature is well mixed with layers of molecular gas, as traced by dark radio features. In this
case, the length of the ionized layer in the N and E arms is 2-3 times larger than the size of the clouds $\sim8''$
($\sim0.3$ pc) associated with the N and E arms.

%The contribution of the continuum emission along
%the path lengths toward and away from  the region of dark clouds 
% results in a very faint dark thermal feature that traces the cool neutral gas.  

%\subsection{The Nature of Bow Shock Stellar Sources} 

%As discussed above, we argue that the Galactic center bow shock stellar sources have stellar disks and that the bow shock 
%arises from the evaporation of disks from YSOs. We have recently discussed this picture .......

\section{Summary and Conclusions}

We have presented five different lines of evidence in favor of on-going star formation near Sgr A*. First, we reported 
the detection of 13 water masers with multiple and single velocity components. After their 
comparison with known OH/IR stars, many of the newly detected water masers  are likely to be 
interstellar tracing star formation 
activity.  The spatial correlation of SiO and HCN (1-0) line emission with water masers provided additional support that 
star formation activity taking  place in the molecular ring. Second, we investigated SED modeling of 64 infrared excess 
sources in the inner pc of Sgr A*. This study indicated that there are massive YSO candidates in this region.
The well-fit YSO models to 19 stellar sources gave us a handle on the total stellar mass $\sim139$ and 
$\sim 119$ \msol\, in Stage I and Stage II, respectively. Using the limited number of YSO candidates, the star 
formation rate within 1 parsec of Sgr A* was estimated to be $\sim1\times10^{-2}$ \msol\, $\rm{yr}^{-1}$. 
We  suggested  the possibility that the YSO candidates  
originate  in the residual outer regions of the 
original gas disk within  which the central cluster of massive stars was formed. 
We argued that stellar bow shock sources in 
this region arise from the interaction of stellar winds from YSO disks with the stream of ionized gas associated with 
the mini-spiral HII region and/or with the UV radiation field from the central cluster. 
Third, the presence of several 
clumps of SiO line emission suggested  tracers of on-going star forming sites in the molecular ring and in the 
immediate vicinity of Sgr A*. We identified a bipolar outflow candidate in one of the clumps of SiO emission in the 
molecular ring. This provided one of the clearest signature of on-going star formation activity in the ring. Our LVG 
analysis of SiO lines indicated a typical molecular density $\sim10^6$ cm$^{-3}$ with an assumed kinetic temperature 
$\sim 150$ K. The total mass swept by protostellar outflow that produced a clump of SiO (5-4) emission 
was estimated to be $\sim$68 \msol. Fourth, 
we presented an EW Ridge of ionized gas with an extent of 20$''$ (0.8 pc). The position velocity diagram of 
the Ne II line 
emission along the length of the EW ionized Ridge showed  a red and blue velocity  component which  is  interpreted as a 
jet arising from a protostar in the molecular ring. Lastly, we presented imprints of molecular gas in 
radio continuum images of the Galactic center. The identified molecular clouds represent a reservoir 
of molecular gas feeding on-going star formation activity within the inner parsec of Sgr A*.

Future high resolution observations of this region will provide us with details 
of massive star formation subject to tidal effects by the nuclear cluster and 
the massive black hole. 
Studies of star formation in the Galactic center region where gravitational potential is dominated by
Sgr A* provide far reaching 
understanding of the mode of  star formation in the nuclei of more active galaxies hosting truly supermassive black holes.

\acknowledgments
We are grateful to S. Gillessen and the MPE group for providing us with  a 3.8$\mu$m image
of the Galactic center.
This work is partially supported by the grant AST-0807400 from the NSF.

\newcommand\refitem{\bibitem[]{}}

% Lai et a.l. 2013; Li et al. 2010

%\bibitem[()]{}
%Beloborodov, A. M., Levin, Y., Eisenhauer, F. et al.  2006, ApJ, 648, 405

\begin{deluxetable}{cccccccccc}
%\label{Catalog}
\tabletypesize{\scriptsize}
\tablecaption{The coordinates and flux density of water masers  in the CMR}
\tablewidth{0pt}
\tablehead{
\colhead{Galactic Coordinates} & \colhead{Name} & \colhead{RA} & \colhead{Dec} & 
\colhead{Peak Velocity$^a$} & 
\colhead{Peak Flux} & \colhead{Cross-Reference} \\
                    &        &   \colhead{1$\sigma$}  & \colhead{$1\sigma$}     &  & \colhead{1$\sigma$}       &}
\startdata
G359.957-0.050 & 1A    & 17 45 42.735     & -28 59 57.390 & 47.29 km s$^{-1}$ & 113.2 mJy & Sjouwerman et al. (2002) \\
               &       & $\pm0.003$       &$\pm0.078$    &      & $\pm2.8$  &\\
  & 1B    &              &               & 54.04       & 100.6     & \\
  & 1C    &              &               & 58.26      & 55.5      & \\
  & 1D    &              &               & 63.33      & 44.9      & \\
 G359.954-0.041  & 2A    & 17 45 40.171 & -28 59 47.818 & 54.04       & 278.1     & \\
               &       & $\pm0.001$       &$\pm0.032$    &      & $\pm2.8$  &\\
  & 2B    &              &  & 56.57      & 62.9      & \\
G359.956-0.035  & 3     & 17 45 39.309 & -28 59 30.127 & 5.07      & 88.0      & \\
                &       & $\pm0.010$       &$\pm0.163$    &     & $\pm2.8$  &\\
G359.935-0.045 & 4     & 17 45 38.623 & -29 00 54.226 & -81.91     & 91.0      & Li et al. (2010)\\
                &       & $\pm0.003$       &$\pm0.098$    &     & $\pm2.8$  &\\
G359.932-0.045 & 5A    & 17 45 38.026 & -29 01 02.634 & 2.53      & 244.2     & \\
                &       & $\pm0.001$       & $\pm0.037$    &      & $\pm2.8$  &\\
  & 5B    &              &               & 6.75      & 22.6      & \\
  & 5C    &              &               & 10.13      & 27.9      & \\
 G359.925-0.053 & 6A    & 17 45 38.954 & -29 01 39.466 & -8.44     & 30.4      & \\
                &       & $\pm0.002$       & $\pm0.086$    &      & $\pm2.8$  &\\
 & 6B    &              &               & -3.37     & 19.8      & \\
  & 6C    &              &               & 0.84     & 39.9      & \\
  & 6D    &              &               & 5.06      & 118.8     & \\
 G359.934-0.050 & 7A    & 17 45 40.148 & -29 00 55.388 & 22.79       & 74.5      & \\
                &       & $\pm0.006$       & $\pm0.146$    &      & $\pm2.8$  &\\
 & 7B    &              &               & 27.02       & 30.2      & \\
  G359.939-0.052 & 8A    & 17 45 40.656 & -29 00 55.679 & 38.84      & 100.2     & \\
                &       & $\pm0.006$       & $\pm0.146$    &      & $\pm2.8$  &\\
& 8B    &              &               & 67.55      & 130.7     & \\
 G359.938-0.064 & 9     & 17 45 43.310 & -29 01 18.880 & -40.53     & 185.7     & \\
                &       & $\pm0.001$       & $\pm0.048$    &      & $\pm2.8$  &\\
G359.943-0.067 & 10     & 17 45 44.747 & -29 01 09.019 & -43.06     & 143.1     & \\
                 &       & $\pm0.003$       &$\pm0.085$    &      & $\pm2.8$  &\\
G359.954-0.053 & 11A    & 17 45 43.022 & -29 00 11.890 & -70.93     & 43.4      & \\
                &       & $\pm0.003$       &$\pm0.131$    &      & $\pm2.8$  &\\
 & 11B    &              &               & -63.33     & 81.6      & \\
 G359.954-0.054 & 12     & 17 45 43.109 & -29 00 13.050 & 32.93      & 104.4     & \\
                &       & $\pm0.003$       &$\pm0.096$    &      & $\pm2.8$  &\\
G359.957-0.050 & 13A    & 17 45 44.326 & -28 59 13.310 & 78.53      & 289.2     & Sjouwerman et al. (2002)\\
                &       & $\pm0.001$       &$\pm0.031$    &      & $\pm2.8$  &\\
  & 13B    &              &               & 80.22      & 319.6     & \\
  & 13C    &              &               & 81.90      & 160.3     & \\
  & 13D    &              &               & 87.81      & 268.0     & \\
  & 13E    &              &               & 91.19      & 99.8      & \\
\enddata
%\tablenotetext{b}{Infrared counterpart 300758} 
\tablenotetext{a}{the uncertainty is 0.84 \kms\, corresponding to the channel width} 
\end{deluxetable}

\begin{deluxetable}{ccrrrrrrrrrrrrrrrr}
\tablecaption{The SED Fitting Results: Physical Parameters of the YSO Candidates}
\tabletypesize{\tiny}
%\tabletypesize{\scriptsize}
\rotate
\tablewidth{0pt}
\tablehead{
\colhead{V06} &
%\colhead{Other} &
\colhead{$\chi_{\rm min}^{2}/$} &
\colhead{n$_{\rm fits}$\tablenotemark{b}} &
\multicolumn{2}{c}{A$_{V}$} &
\multicolumn{2}{c}{L$_{\star}$ (10$^{4}$ L$_{\odot}$)} &
\multicolumn{2}{c}{T$_{\star}$ (10$^{4}$ K} &
\multicolumn{2}{c}{M$_{\star}$(M$_{\odot}$)} &
\multicolumn{2}{c}{M$_{env}$(M$_{\odot}$)} &
\multicolumn{2}{c}{M$_{disk}$(M$_{\odot}$)} &
\multicolumn{2}{c}{Evol. Stage} &
\colhead{\#}
 \\

\colhead{V0 ID} &
\colhead{$pt$\tablenotemark{a}} &
\colhead{} &
\colhead{best} &
\colhead{ave\tablenotemark{c}} &
\colhead{best} &
\colhead{ave} &
\colhead{best} &
\colhead{ave} &
\colhead{best} &
\colhead{ave} &
\colhead{best} &
\colhead{ave} &
\colhead{best} &
\colhead{ave} &
\colhead{best} &
\colhead{ave} &
\colhead{data}
}
\startdata
      5 &     1.36 &     367 &  25.0 &  25.3 (1.0) &   2.4 &   1.1 (0.05)  &   3.21 &   2.74 (0.02) &   15.9 &     11.7 (0.2) &     0.00 &     0.15 (0.07)   &    0.000 &    0.006 (0.001) &    III &     II &   6 \\
      6 &      1.33 &      31 &  25.0 &  25.0 (0.0) &   7.1 &   6.0 (0.27) &   3.76 &   3.66 (0.03) &   23.7 &     22.1 (0.4) &     0.00 &      0.0 (0.0)    &    0.043 &    0.130 (0.027) &     II &     II &   9 \\
      10&     0.53 &    5700 &  25.5 &  27.2 (2.4) &    0.12 &    0.33 (0.01)   &   1.93 &   1.94 (0.01)  &    6.4 &    7.85 (0.03) &     9.2  &     22.1 (1.8)    &    0.056 &    0.033 (0.001) &     II &      I &   4 \\
      11&       0.03 &    4583 &  29.8 &  27.7 (2.4) &    0.32 &    0.37 (0.01)   &    0.51 &   1.88 (0.01) &   10.6 &    8.55 (0.03) &     1293 &     52.3 (3.3)    &    0.219 &    0.054 (0.002) &      I &      I &   5 \\
      12&     0.03 &    3039 &  29.2 &  28.3 (2.2) &    0.49 &    0.46 (0.01)   &    0.41 &   1.98 (0.01) &   11.9 &    9.21 (0.04) &     1415 &     65.7 (4.8)    &    0.000 &    0.061 (0.002) &      I &      I &   5 \\
      13&      0.43 &      29 &  25.0 &  25.1 (0.6) &   7.09 &   6.58 (0.51) &   3.77 &   3.45 (0.05) &   23.7 &     22.8 (0.8) &     0.00 &       160 (114)   &    0.043 &    0.137 (0.028) &     II &      I &   7 \\
      14&     0.27 &     627 &  29.1 &  26.9 (2.2) &   2.43 &   1.77 (0.04)  &   3.21 &   2.98 (0.01) &   15.9 &     13.8 (0.1) &     0.00 &      2.7 (1.5)    &    0.010 &    0.029 (0.003) &     II &      I &   7 \\
      15&      0.19 &    2192 &  29.9 &  26.8 (2.2) &   1.15 &    0.87 (0.01)  &   2.84 &   2.43 (0.01) &   12.2 &     11.2 (0.1) &     0.00 &     83.0 (7.2)    &    0.004 &    0.060 (0.003) &     II &      I &   7 \\
      18&      0.49 &    1245 &  25.0 &  26.7 (2.2) &    0.94 &   1.28 (0.02)  &   2.75 &   2.75 (0.01) &   11.4 &     12.2 (0.1) &     0.00 &     21.8 (4.9)    &    0.066 &    0.042 (0.003) &     II &      I &   7 \\
      19&    0.06 &    5145 &  25.0 &  27.0 (2.3) &   1.17 &    0.93 (0.01)   &   2.85 &   2.41 (0.01) &   12.3 &   11.39 (0.03) &     20.4 &     81.5 (3.2)    &    0.000 &    0.067 (0.002) &     II &      I &   7 \\
      26&     0.24 &    4954 &  25.0 &  27.6 (2.4) &    0.21 &    0.17 (0.003)   &    0.43 &   1.66 (0.01)  &   10.1 &    6.40 (0.03) &     33.7 &     13.8 (1.5)    &    0.040 &    0.026 (0.001) &      I &      I &   5 \\
      28&     1.85 &     776 &  25.0 &  25.1 (0.6) &   1.21 &   1.61 (0.03)  &   2.81 &   2.92 (0.01) &   12.0 &     13.2 (0.1) &     0.00 &      0.0 (0.0)    &    0.038 &    0.016 (0.002) &     II &     II &   8 \\
      31&     1.98 &     193 &  25.9 &  26.3 (2.0) &   2.76 &   3.16 (0.08)  &   3.27 &   3.31 (0.01) &   16.6 &     17.2 (0.2) &     0.00 &      0.0 (0.0)    &    0.000 &    0.034 (0.005) &     II &     II &   7 \\
      32&     2.37 &     547 &  25.0 &  25.3 (1.0) &   2.68 &   1.96 (0.03)  &   3.25 &   3.00 (0.02) &   16.4 &     14.6 (0.1) &     0.00 &     24.4 (3.7)    &    0.018 &    0.037 (0.003) &     II &      I &   9 \\
      39&     0.80 &     879 &  30.0 &  27.8 (2.3) &   1.62 &   1.33 (0.03)  &    0.41 &   2.57 (0.03) &   23.9 &     12.7 (0.1) &      408 &     87.8 (13.1)   &    0.221 &    0.056 (0.004) &      I &      I &   5 \\
      47&    1.32 &     186 &  30.0 &  27.8 (2.0) &   4.62 &   2.84 (0.09)  &   3.52 &   3.23 (0.02) &   20.0 &     16.4 (0.2) &     0.00 &      0.0 (0.0)    &    0.068 &    0.016 (0.003) &     II &     II &   9 \\
      48&    0.36 &     565 &  30.0 &  27.5 (2.2) &   3.86 &   1.98 (0.04)  &   3.39 &   3.04 (0.01) &   18.1 &     14.3 (0.1) &     0.00 &      0.7 (0.7)    &    0.052 &    0.027 (0.003) &     II &     II &   8 \\
      54&    0.48 &      19 &  25.0 &  25.0 (0.0) &   7.09 &   6.63 (0.37) &   3.76 &   3.70 (0.04) &   23.7 &     22.9 (0.6) &     0.00 &      0.0 (0.0)    &    0.043 &    0.176 (0.030) &     II &     II &   9 \\
      57&    0.77 &    2549 &  30.0 &  27.3 (2.2) &    0.40 &   0.55 (0.01)  &   2.38 &   2.35 (0.082)  &    8.8 &    9.19 (0.04) &     0.00 &     14.6 (1.8)    &    0.007 &    0.027 (0.001) &     II &      I &   6 \\
\enddata
\tablenotetext{a}{$\chi_{\rm min}^{2}/pt$ is a $\chi^{2}$ per data point for the best-fit model.}
\tablenotetext{b}{~n$_{\rm fits}$ is a number of fits with $\chi^{2}/pt$ between $\chi^{2}_{\rm min}/pt$ and $\chi^{2}_{\rm min}/pt+$2.}
\tablenotetext{c}{The uncertainties represent standard deviations of the mean.} 
\end{deluxetable}

\begin{deluxetable}{cccccccc}
%\label{Catalog}
\tabletypesize{\scriptsize}
\tablecaption{YSO Candidate Positions and Literature References}
\tablewidth{0pt}
\tablehead{
\colhead{V06 ID} & \colhead{Other Name} & \colhead{RA (J2000)} & \colhead{Dec (J2000)}  & \colhead{Cross-Reference$^a$} \\
%                    &        &   \colhead{1$\sigma$}  & \colhead{$1\sigma$}     &  & \colhead{1$\sigma$}       &
}
\startdata
1  & IRS 1W & 17 45 40.44 & -29 00 27.7 & 1, 3, 6, 8 \\
5  & \nodata & 17 45 40.46 &  -29 00 29.2 & \\
6  & IRS 2L & 17 45 39.78 &  -29 00 32.1 & 8, 9\\
10 & \nodata & 17 45 39.98 & -29 00 24.3 & \\
11 & IRS 4  & 17 45 40.84 &-29 00 34.2 & \\
12 & \nodata & 17 45 40.80 & -29 00 37.0 & \\
13 & IRS 5  & 17 45 40.70 & -29 00 18.5 & 1, 4, 5, 6, 8\\
14 & IRS 5NE & 17 45 41.02 & -29 00 17.7 & 5, 8\\
15 & IRS 5E & 17 45 40.87 & -29 00 19.0 & 5\\
18 & IRS 5S & 17 45 40.74 &-29 00 20.3 & 5\\
19 & IRS 5SE1 & 17 45 40.86 &-29 00 21.3 &5 \\
26 & \nodata  & 17 45 40.03 &-29 00 20.4 & \\
28 & \nodata & 17 45 40.33 &-29 00 21.1 & \\
31 & \nodata & 17 45 40.32 & -29 00 33.6 & \\
32 & \nodata & 17 45 40.36 & -29 00 35.6 & \\
35 & IRS 10W & 17 45 40.54 & -29 00 23.1 & 1, 2, 6, 8\\
39 & \nodata & 17 45 40.52 & -29 00 31.1 & \\
47 & IRS 13N & 17 45 39.81 &-29 00 29.2 &7, 10 \\
48 & IRS 13NE & 17 45 39.84 &-29 00 29.7 & 7\\
54 & IRS 21 & 17 45 40.22 & -29 00 31.0 & 1, 2, 4, 8, 11\\
57 & IRS 34SW & 17 45 39.74 & -29 00 26.7 & \\
\enddata
\tablenotetext{a}{References: (1) Sanchez-Bermudez et al. 2014; (2) Yoshikawa et al. 2014; 
(3), Paumard et al. 2006; (4) Tanner et al. 2002; (5) Perger et al. 2007; (6) Tanner et al. 2005; (7) Eckart et al. (2013); 
(8) Buchholz et al. 2013; (9) Moultaka et al. (2009); (10) Muzic et al. 2010; (11) Yusef-Zadeh et al. 2014  
} 
\end{deluxetable}

%\tablenotetext{a}{the uncertainty is 0.84 \kms\, corresponding to the channel width} 

\begin{deluxetable}{rcccccc}
\label{Catalog}
\tablecaption{Gaussian Line Parameters of Fitted SiO (5-4) Sources}
\tabletypesize{\scriptsize}
\tablewidth{0pt}
%\rotate
\tablehead{
\colhead{Source} & \colhead{RA} & \colhead{DEC} & \colhead{Peak Intensity} & \colhead{Center Velocity} & \colhead{FWHM} &
 \colhead{FWZI}\\ &
\colhead{(J2000)} & \colhead{(J2000)} & \colhead{mJy beam$^{-1}$} & \colhead{km s$^{-1}$} & \colhead{km s$^{-1}$} & \colhead{km
s$^{-1}$}
}
\startdata
10 & 17:45:40.31 & -29 00 43.77 & 174.7 $\pm$ 20.0 & -58.88 $\pm$ 1.87 & 33.43 $\pm$ 4.41 & 68 $\pm$ 20\\
10 (ALMA)&             &              & 71.32 $\pm$ 2.75 & -42.98 $\pm$ 0.39 & 21.02 $\pm$ 0.95 & 56 $\pm$ 18\\
12 & 17:45:41.08 & -29 00 37.46 & 91.58 $\pm$ 19.3 & -10.60 $\pm$ 5.34 & 51.68 $\pm$ 12.6 & 96 $\pm$ 15\\
13 & 17:45:40.69 & -29 00 45.83 & 238.9 $\pm$ 28.2 & -36.61 $\pm$ 0.96 & 17.65 $\pm$ 2.52 & 35 $\pm$ 10\\
14 & 17:45:40.37 & -29 00 53.73 & 263.2 $\pm$ 19.0 & -54.42 $\pm$ 0.99 & 28.11 $\pm$ 2.34 & 49 $\pm$ 14\\
15 & 17:45:40.20 & -29 01 00.20 & 557.9 $\pm$ 34.9 & -74.76 $\pm$ 1.06 & 34.67 $\pm$ 2.50 & 159 $\pm$ 40\\
16 & 17:45:39.78 & -29 01 05.94 & 450.1 $\pm$ 26.3 & -76.72 $\pm$ 0.98 & 34.14 $\pm$ 2.30 & 62 $\pm$ 8\\
17 & 17:45:41.32 & -29 01 01.63 & 317.3 $\pm$ 32.1 &  10.83 $\pm$ 1.11 & 12.71 $\pm$ 1.56 & 27 $\pm$ 5\\
18 & 17:45:41.55 & -29 00 57.08 & 312.9 $\pm$ 16.9 &  9.424 $\pm$ 0.80 & 10.86 $\pm$ 0.92 & 14 $\pm$ 10\\
19 & 17:45:41.84 & -29 00 57.56 & 317.4 $\pm$ 10.6 &  11.12 $\pm$ 4.09 & 9.842 $\pm$ 4.94 & 14 $\pm$ 8\\
20 & 17:45:44.85 & -29 00 25.25 & 282.3 $\pm$ 31.0 &  36.61 $\pm$ 0.89 & 17.70 $\pm$ 2.35 & 27 $\pm$ 5\\
21 & 17:45:43.73 & -29 00 16.63 & 327.0 $\pm$ 57.9 &  35.85 $\pm$ 0.88 & 15.09 $\pm$ 3.44 & 23 $\pm$ 5\\
22 & 17:45:43.52 & -29 00 07.06 & 524.3 $\pm$ 26.9 &  43.76 $\pm$ 0.63 & 24.97 $\pm$ 1.48 & 46 $\pm$ 12\\
23 & 17:45:43.74 & -28 59 59.88 & 341.8 $\pm$ 39.5 &  39.34 $\pm$ 1.61 & 28.56 $\pm$ 3.81 & 55 $\pm$ 10\\
24 & 17:45:41.75 & -28 59 47.62 & 316.1 $\pm$ 34.2 &  87.37 $\pm$ 0.96 & 18.11 $\pm$ 2.28 & 34 $\pm$ 8\\
25 & 17:45:39.49 & -28 59 51.02 & 268.6 $\pm$ 30.6 &  10.17 $\pm$ 1.90 & 34.06 $\pm$ 4.48 & 65 $\pm$ 20\\
26 & 17:45:39.31 & -29 00 02.95 & 182.8 $\pm$ 24.2 &  28.15 $\pm$ 1.71 & 26.37 $\pm$ 4.04 & 51 $\pm$ 15\\
27 & 17:45:39.14 & -29 00 16.74 & 227.5 $\pm$ 36.8 &  37.35 $\pm$ 1.68 & 21.22 $\pm$ 3.97 & 40 $\pm$ 13\\
28 & 17:45:38.17 & -29 00 33.00 & 225.2 $\pm$ 20.8 &  12.76 $\pm$ 0.93 & 20.63 $\pm$ 2.21 & 39 $\pm$ 10\\
29 & 17:45:38.46 & -29 00 33.68 & 243.3 $\pm$ 10.5 &  61.78 $\pm$ 1.52 & 25.25 $\pm$ 3.24 & 45 $\pm$ 20\\
30 & 17:45:37.44 & -29 00 44.36 & 153.1 $\pm$ 21.8 & -23.25 $\pm$ 2.76 & 39.52 $\pm$ 6.59 & 56 $\pm$ 15\\
31 & 17:45:38.25 & -29 00 41.49 & 505.5 $\pm$ 34.8 & -29.03 $\pm$ 1.00 & 29.60 $\pm$ 2.35 & 58 $\pm$ 20\\
32 & 17:45:38.37 & -29 00 48.61 & 575.9 $\pm$ 60.1 & -39.59 $\pm$ 2.58 & 50.49 $\pm$ 6.09 & 92 $\pm$ 25\\
33 & 17:45:37.97 & -29 00 48.47 & 597.3 $\pm$ 34.2 & -46.93 $\pm$ 1.50 & 53.55 $\pm$ 3.54 & 108 $\pm$ 25\\
34 & 17:45:38.04 & -29 00 54.08 & 479.0 $\pm$ 32.2 & -39.21 $\pm$ 1.34 & 40.76 $\pm$ 3.17 & 84 $\pm$ 18\\
35 & 17:45:38.46 & -29 01 02.71 & 611.5 $\pm$ 3.63 & -124.3 $\pm$ 0.12 & 36.57 $\pm$ 0.31 & 80 $\pm$ 50\\
36 & 17:45:39.04 & -29 01 03.39 & 206.6 $\pm$ 18.6 & -100.5 $\pm$ 1.08 & 24.46 $\pm$ 2.54 & 51 $\pm$ 15\\
37 & 17:45:35.88 & -29 01 08.01 & 178.1 $\pm$ 31.2 & -40.70 $\pm$ 2.32 & 27.01 $\pm$ 5.46 & 41 $\pm$ 20\\
\enddata
\end{deluxetable}

\begin{deluxetable}{rcccccc}
%\rotate
\tablecaption{Gaussian Line Parameters of Fitted SiO (5-4) and (2-1) Sources}
%\tablecolumns{7}
\tabletypesize{\scriptsize}
\tablewidth{0pt}
\tablehead{ 
    \colhead{}  & \multicolumn{3}{c}{\underline{CARMA}} & \multicolumn{3}{c}{\underline{SMA}} \\
    \colhead{}  & \colhead{Peak}            & \colhead{Center}      & \colhead{FWHM}
                & \colhead{Peak}            & \colhead{Center}      & \colhead{FWHM} \\
    \colhead{}  & \colhead{mJy beam$^{-1}$} & \colhead{km s$^{-1}$} & \colhead{km s$^{-1}$}
                & \colhead{mJy beam$^{-1}$} & \colhead{km s$^{-1}$} & \colhead{km s$^{-1}$}
}
\startdata
10 &  45.2$\pm$64.2 &  -40.8$\pm$3.2 & 46.8$\pm$ 7.6 & 167.2$\pm$ 18.3 &  -41.8$\pm$1.3 & 24.2$\pm$ 3.0 \\
12 &                  &                  &                 & 218.8$\pm$ 10.1 & -9.4$\pm$1.7 & 74.4$\pm$ 4.0 \\
13 &  87.3$\pm$13.2 &  -19.2$\pm$3.2 & 61.1$\pm$11.0 & 353.2$\pm$ 23.6 &  -18.2$\pm$0.7 & 23.7$\pm$ 1.8 \\
14 & 116.0$\pm$ 8.7 &  -38.4$\pm$1.1 & 32.0$\pm$ 2.7 & 352.9$\pm$ 20.9 &  -35.8$\pm$0.9 & 31.0$\pm$ 2.1 \\
15 & 131.7$\pm$ 9.6 &  -50.8$\pm$1.5 & 43.1$\pm$ 3.6 & 488.9$\pm$ 23.8 &  -49.0$\pm$1.0 & 42.2$\pm$ 2.3 \\
16 & 162.0$\pm$ 9.9 &  -60.0$\pm$0.7 & 26.4$\pm$ 1.8 & 461.8$\pm$ 39.5 &  -58.4$\pm$1.1 & 26.3$\pm$ 2.6 \\
17 & 20.2$\pm$12.7 &   27.2$\pm$0.6 & 21.4$\pm$ 1.5 & 420.7$\pm$ 44.6 &   30.5$\pm$0.6 & 12.0$\pm$ 1.4 \\
18 & 230.9$\pm$12.1 &   30.7$\pm$0.4 & 16.4$\pm$ 1.0 & 395.6$\pm$ 20.9 &   30.7$\pm$0.3 & 13.3$\pm$ 0.8 \\
19 & 244.7$\pm$12.7 &   28.0$\pm$0.2 & 10.4$\pm$ 0.6 & 403.4$\pm$ 46.2 &   28.5$\pm$0.5 & 10.1$\pm$ 1.3 \\
20 &  83.6$\pm$15.9 &   37.8$\pm$0.9 & 10.3$\pm$ 2.2 & 158.0$\pm$ 198.4 &   33.4$\pm$3.4 &  6.6$\pm$10.1 \\
21 & 222.2$\pm$11.3 &   57.1$\pm$0.5 & 23.3$\pm$ 1.4 & 532.1$\pm$ 54.5  &   54.11$\pm$0.7 & 14.6$\pm$ 1.7 \\
22 & 238.6$\pm$ 9.7 &   62.7$\pm$0.5 & 26.7$\pm$ 1.3 & 631.2$\pm$ 39.5 &   61.8$\pm$0.7 & 25.4$\pm$ 1.8 \\
23 & 222.9$\pm$ 7.9 &   63.7$\pm$0.5 & 31.3$\pm$ 1.3 & 683.7$\pm$ 41.1 &   59.1$\pm$0.7 & 25.8$\pm$ 1.8 \\
24 & 120.1$\pm$10.1 &  104.0$\pm$0.8 & 20.3$\pm$ 1.9 & 254.2$\pm$ 22.2 &  107.1$\pm$0.8 & 20.4$\pm$ 2.0 \\
25 &  71.5$\pm$ 5.4 &   38.2$\pm$2.0 & 54.5$\pm$ 4.7 & 320.5$\pm$ 17.1 &   32.0$\pm$1.1 & 43.1$\pm$ 2.6 \\
26 &  70.7$\pm$ 9.7 &   48.4$\pm$1.8 & 26.6$\pm$ 4.2 & 137.9$\pm$ 21.0 &   42.5$\pm$1.9 & 25.6$\pm$ 4.5 \\
27 &  77.6$\pm$ 8.4 &   61.9$\pm$1.5 & 28.2$\pm$ 3.5 & 265.2$\pm$ 25.1 &   60.4$\pm$1.1 & 24.1$\pm$ 2.6 \\
   &               &               &                       & 14.86$\pm$ 11.2 &  -13.9$\pm$1.9 & 50.8$\pm$ 4.6 \\
28 &  68.9$\pm$ 5.9 &   34.1$\pm$0.8 & 20.2$\pm$ 2.0 & 190.6$\pm$ 22.9 &   35.9$\pm$1.4 & 24.9$\pm$ 3.4 \\
   &  49.6$\pm$ 8.4 &   87.5$\pm$2.0 & 24.3$\pm$ 4.8 & 111.6$\pm$ 17.9 &   87.8$\pm$1.7 & 22.4$\pm$ 4.2 \\
29 & 128.1$\pm$ 5.9 &   82.8$\pm$0.5 & 23.5$\pm$ 1.2 & 307.1$\pm$ 37.6 &   84.8$\pm$1.3 & 21.5$\pm$ 3.2 \\
30 &  36.3$\pm$ 6.4 &  -11.5$\pm$4.4 & 51.5$\pm$10.5 &                   &                  &                 \\
31 & 199.1$\pm$ 6.4 &  -11.7$\pm$0.6 & 37.3$\pm$ 1.4 & 613.3$\pm$ 25.1 &  -10.8$\pm$0.6 & 34.4$\pm$ 1.6 \\
32 & 228.7$\pm$ 9.5 &  -22.6$\pm$1.0 & 53.5$\pm$ 2.5 & 730.6$\pm$ 26.4 &  -19.5$\pm$0.9 & 54.4$\pm$ 2.2 \\
33 & 250.5$\pm$ 8.7 &  -22.4$\pm$0.9 & 56.4$\pm$ 2.2 & 831.9$\pm$ 21.8 &  -23.1$\pm$0.7 & 58.5$\pm$ 1.7 \\
34 & 211.8$\pm$ 9.2 &  -23.3$\pm$1.1 & 52.3$\pm$ 2.6 & 807.3$\pm$ 21.8 &  -22.3$\pm$0.6 & 50.8$\pm$ 1.5 \\
35 & 152.9$\pm$10.6 & -104.6$\pm$1.1 & 34.6$\pm$ 2.7 & 620.2$\pm$ 15.7 & -103.4$\pm$0.4 & 37.5$\pm$ 1.1 \\
36 &  64.5$\pm$11.9 &  -81.7$\pm$1.1 & 12.5$\pm$ 2.6 & 183.6$\pm$ 27.2 &  -79.3$\pm$1.4 & 20.1$\pm$ 3.4 \\
   &               &               &                       & 116.5$\pm$ 34.1 &   -4.6$\pm$3.1 & 2.20$\pm$ 7.5 \\
37 & 107.4$\pm$11.6 &  -19.4$\pm$1.3 & 24.9$\pm$ 3.2 & 444.4$\pm$ 42.9 &  -17.2$\pm$1.1 & 24.4$\pm$ 2.7
\enddata
\vspace{-2cm}
\end{deluxetable}

%\vfill\eject

\begin{figure}
\figurenum{1}
\centering
\includegraphics[scale=0.8,angle=0]{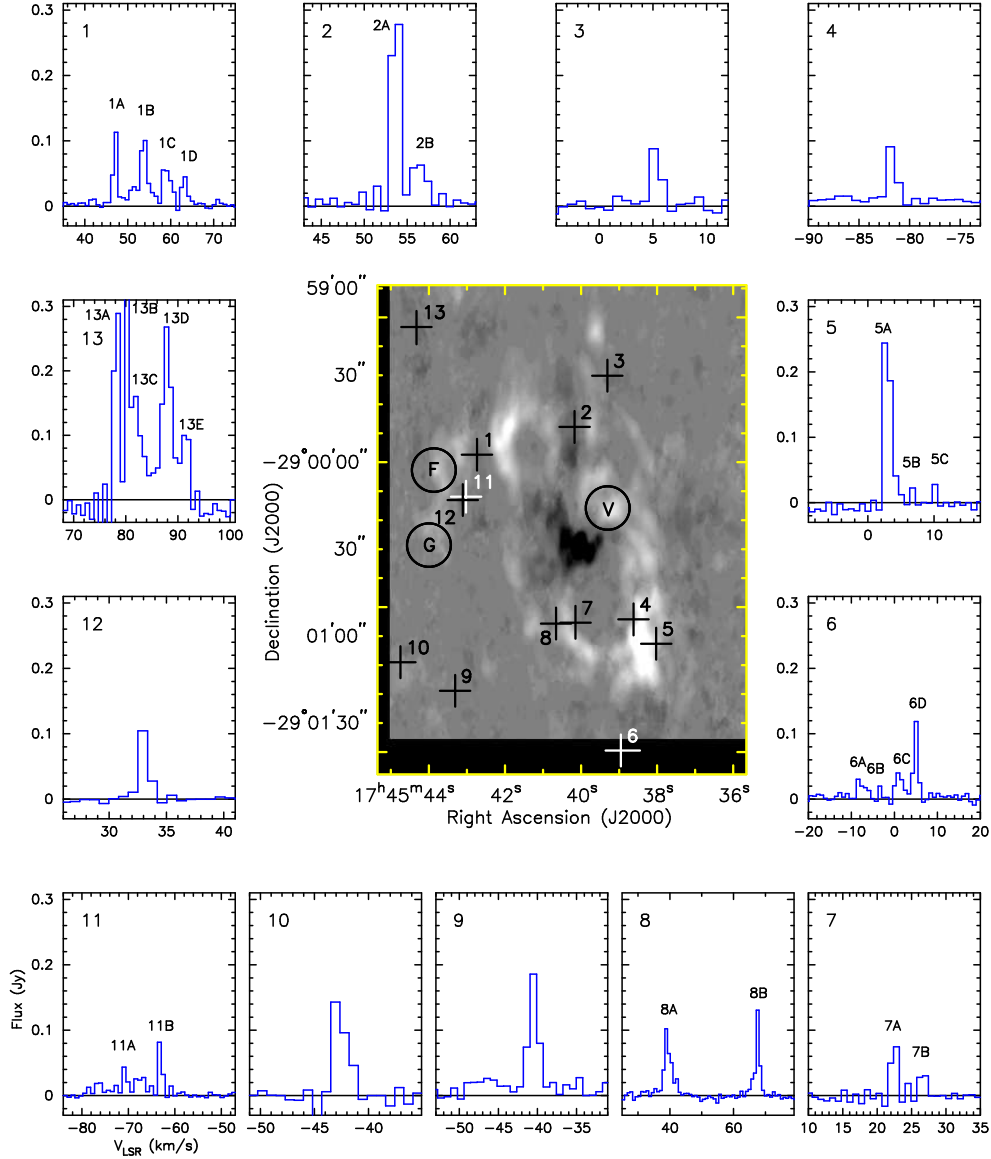}
\caption{%{\it (a)} 
The positions of water masers (this work) and collisionally excited methanol masers (YBWR08) 
are drawn as crosses 
and circles, respectively, on an HCN(1-0) map (Christopher et al. 2005). 
The size of the circles correspond to the spatial  resolution of 
Green Bank Telescope (GBT) observations (FWHM$\sim15''$; Yusef-Zadeh et al. 2008). 
The surrounding plots show the spectra of  13 water masers.}
\end{figure}  

\begin{figure}
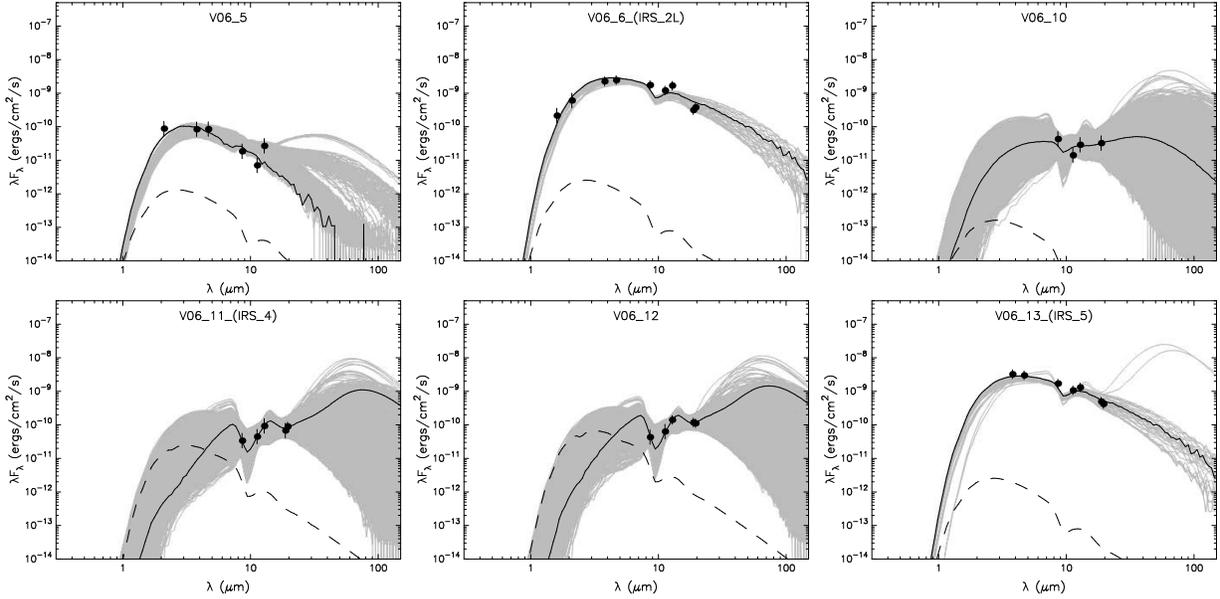

\figurenum{2}
\centering
\includegraphics[width=0.32\textwidth]{f2a_V06_5.eps}
\includegraphics[width=0.32\textwidth]{f2b_V06_6__IRS_2L_.eps}
\includegraphics[width=0.32\textwidth]{f2c_V06_10.eps}\\
\includegraphics[width=0.32\textwidth]{f2d_V06_11__IRS_4_.eps}
\includegraphics[width=0.32\textwidth]{f2e_V06_12.eps}
\includegraphics[width=0.32\textwidth]{f2f_V06_13__IRS_5_.eps}
\caption{
SEDs and YSO model fits for 19 YSO candidates well-fit with YSO models. 
The sources are located in the central YSO 
cluster near Sgr A*. Filled circles are flux values from the Viehmann et al. (2006) VLT/VISIR observations (1.6 --19.5 
$\mu$m). The flux error bars are plotted if they are larger than the data points. The solid black line in each plot shows the 
best fit YSO model (Robitaille et al. 2006). The acceptable YSO fits are shown in gray. The dashed line is the central 
stellar atmosphere corresponding to the best fit model, extincted by the fitted foreground extinction. 
The physical parameters estimated based on the YSO SED fitting are listed in Table 2. 
The small  number on each plot 
identifies 
The source number given in Table 2 of Viehmann et al. (2006) followed by its near-IR identified name 
 is shown  on each plot.}
\end{figure}  

\begin{figure}
\figurenum{2b}
\centering
\includegraphics[width=0.32\textwidth]{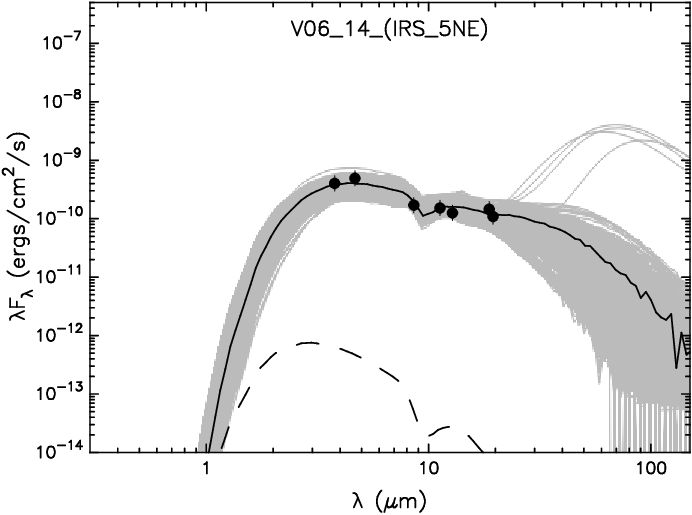}
\includegraphics[width=0.32\textwidth]{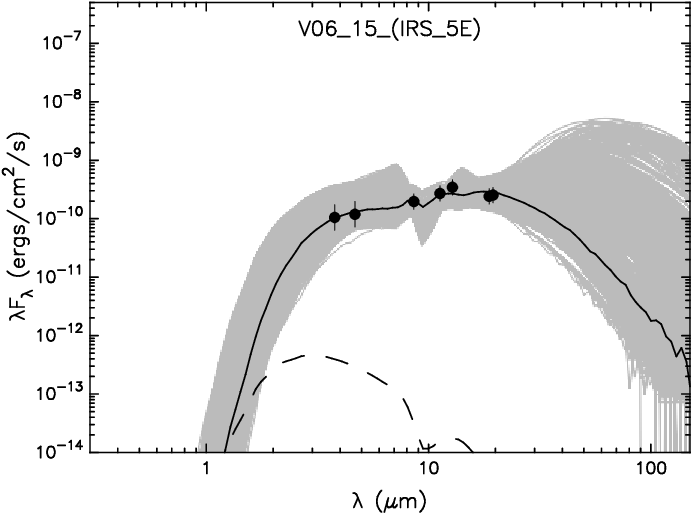}
\includegraphics[width=0.32\textwidth]{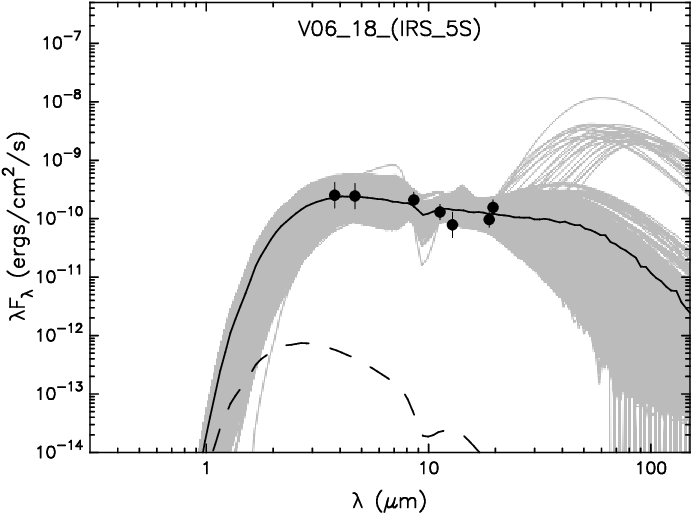}\\
\includegraphics[width=0.32\textwidth]{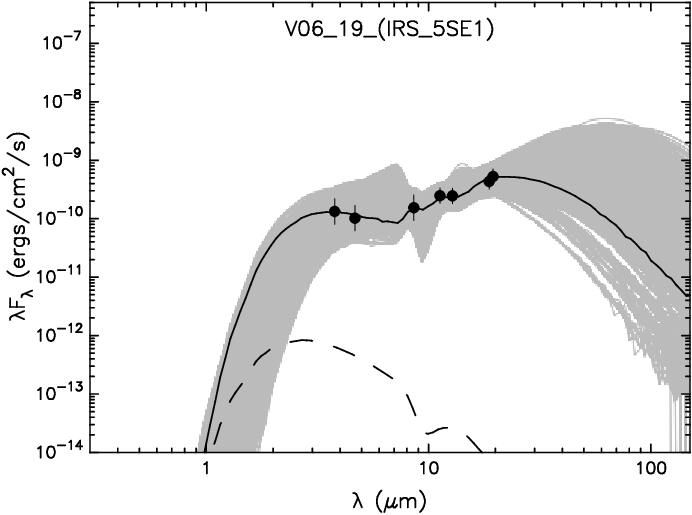}
\includegraphics[width=0.32\textwidth]{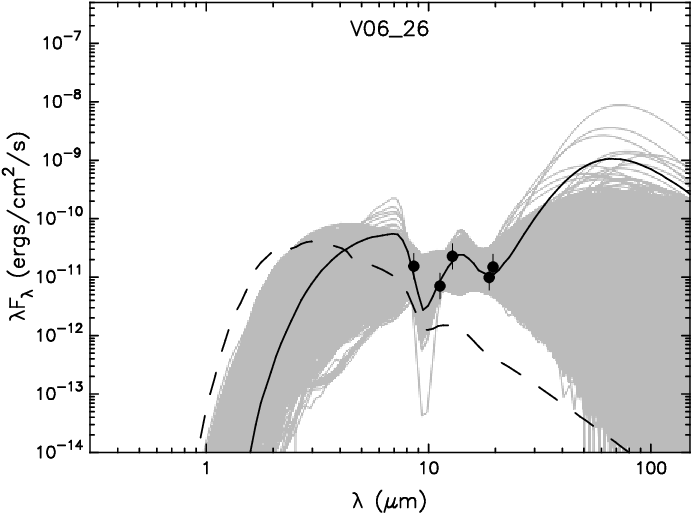}
\includegraphics[width=0.32\textwidth]{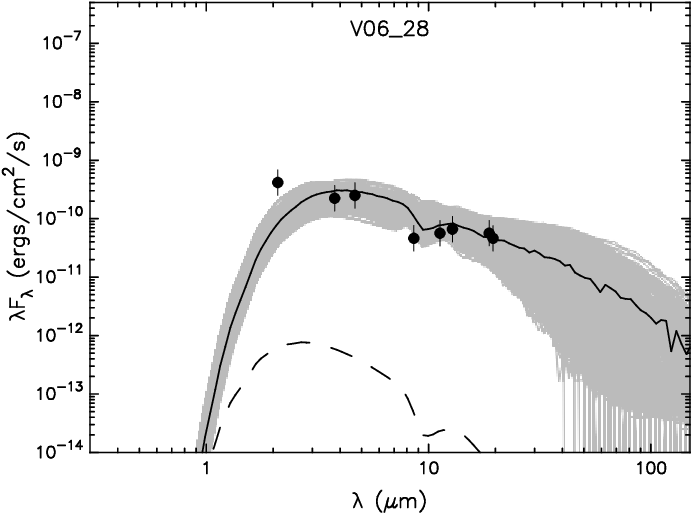}
\end{figure}  

\begin{figure}
\figurenum{2c}
\centering
\includegraphics[width=0.32\textwidth]{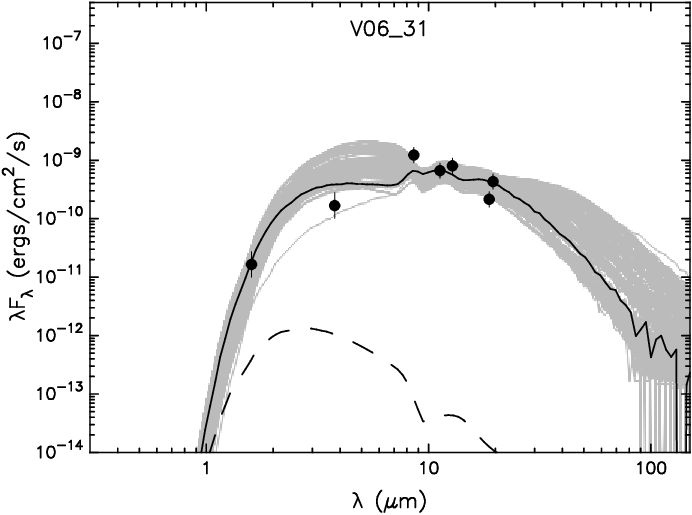}
\includegraphics[width=0.32\textwidth]{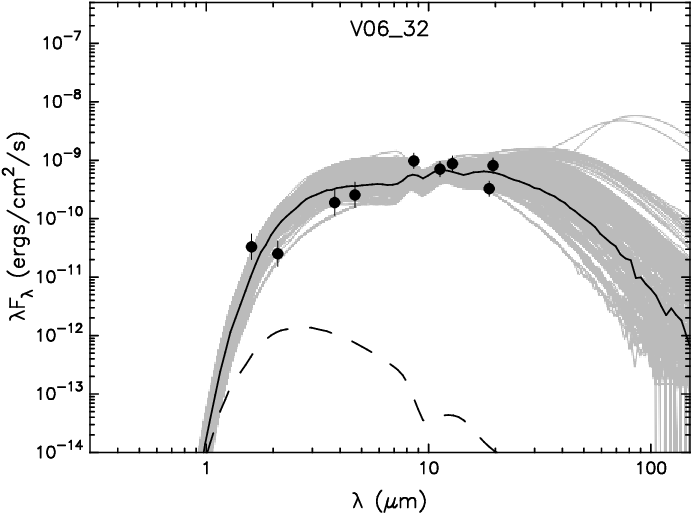}
\includegraphics[width=0.32\textwidth]{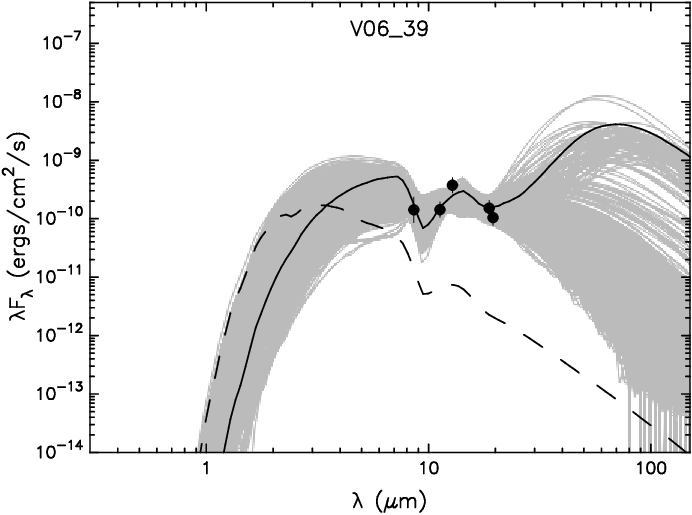}\\
\includegraphics[width=0.32\textwidth]{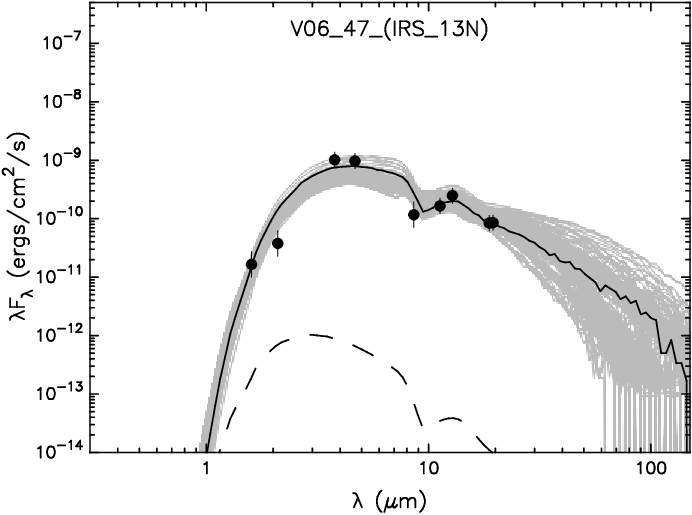}
\includegraphics[width=0.32\textwidth]{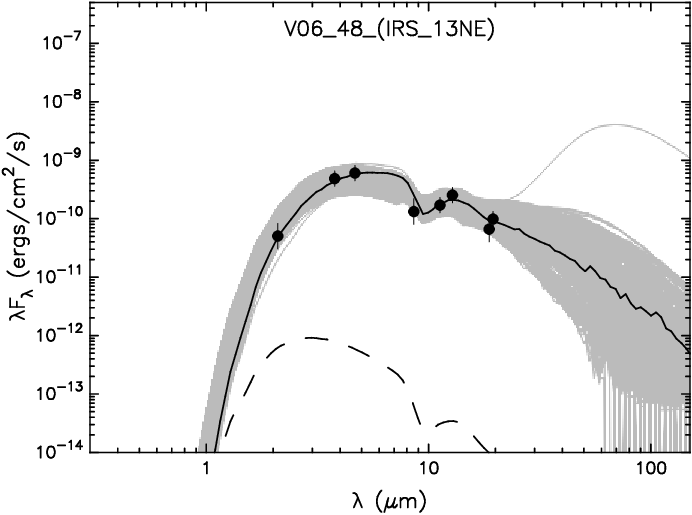}
\includegraphics[width=0.32\textwidth]{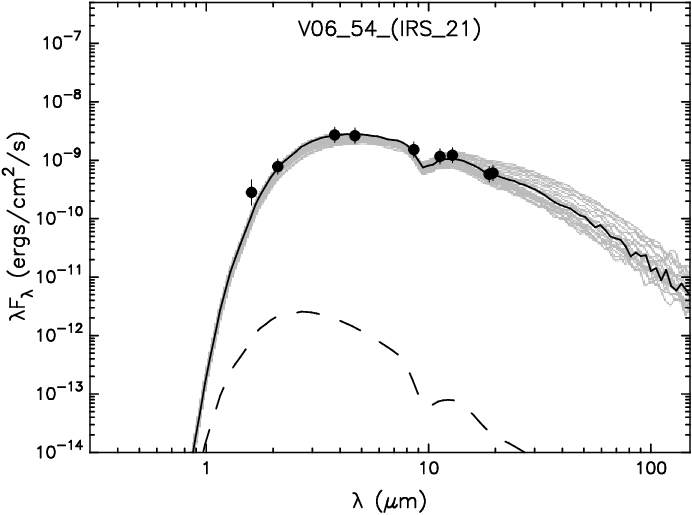}
\includegraphics[width=0.32\textwidth]{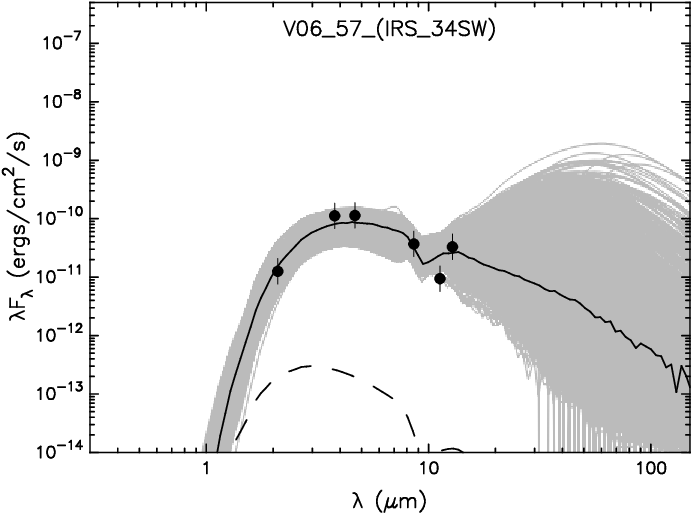}
\end{figure}  

%The SED fitted plots of 35 candidate YSOs %found in the central YSO cluster near Sgr A*. The points represent observed 
%data data points taken at different wavelengths (Viehmann et al. 2006). The shaded line show family of fits to the 
%data. Dashed lines represent models of the photosphere. The parameters of the fits are shown in Table 2.

\begin{figure}
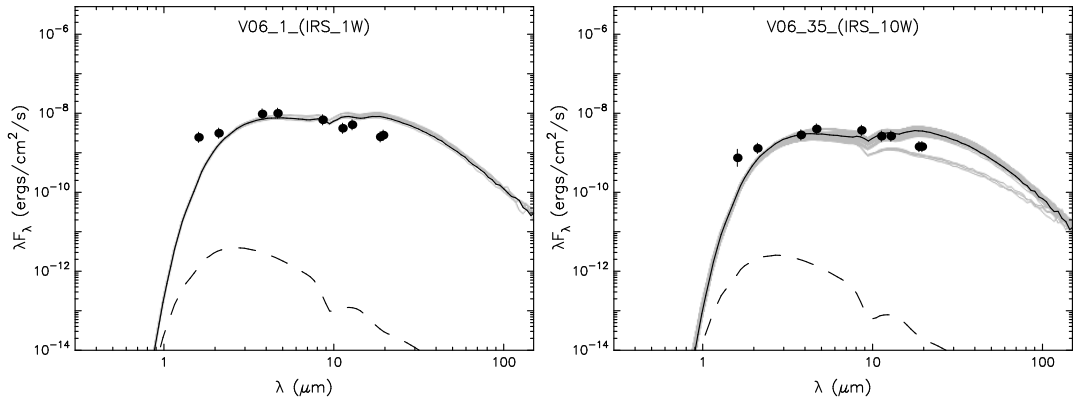

\figurenum{3}
\center
\includegraphics[scale=0.6,angle=0]{fig3a_v06_IRS1W.eps}
\includegraphics[scale=0.6,angle=0]{fig3b_v06_IRS10W.eps}
\caption{
%{\it (a) Top }
Similar to Figure 2 except that 
SEDs YSO model fits to IRS 1W and IRS 10W, as 
 presented in the left and right panels, 
give   $\chi^2/pt$ that are  12 and 5, respectively.  
The YSO fitted SEDs  to these sources 
are more uncertain than those shown in Figure 2. 
}
\end{figure}

\begin{figure}
\figurenum{4}
\center
\includegraphics[scale=0.5,angle=0]{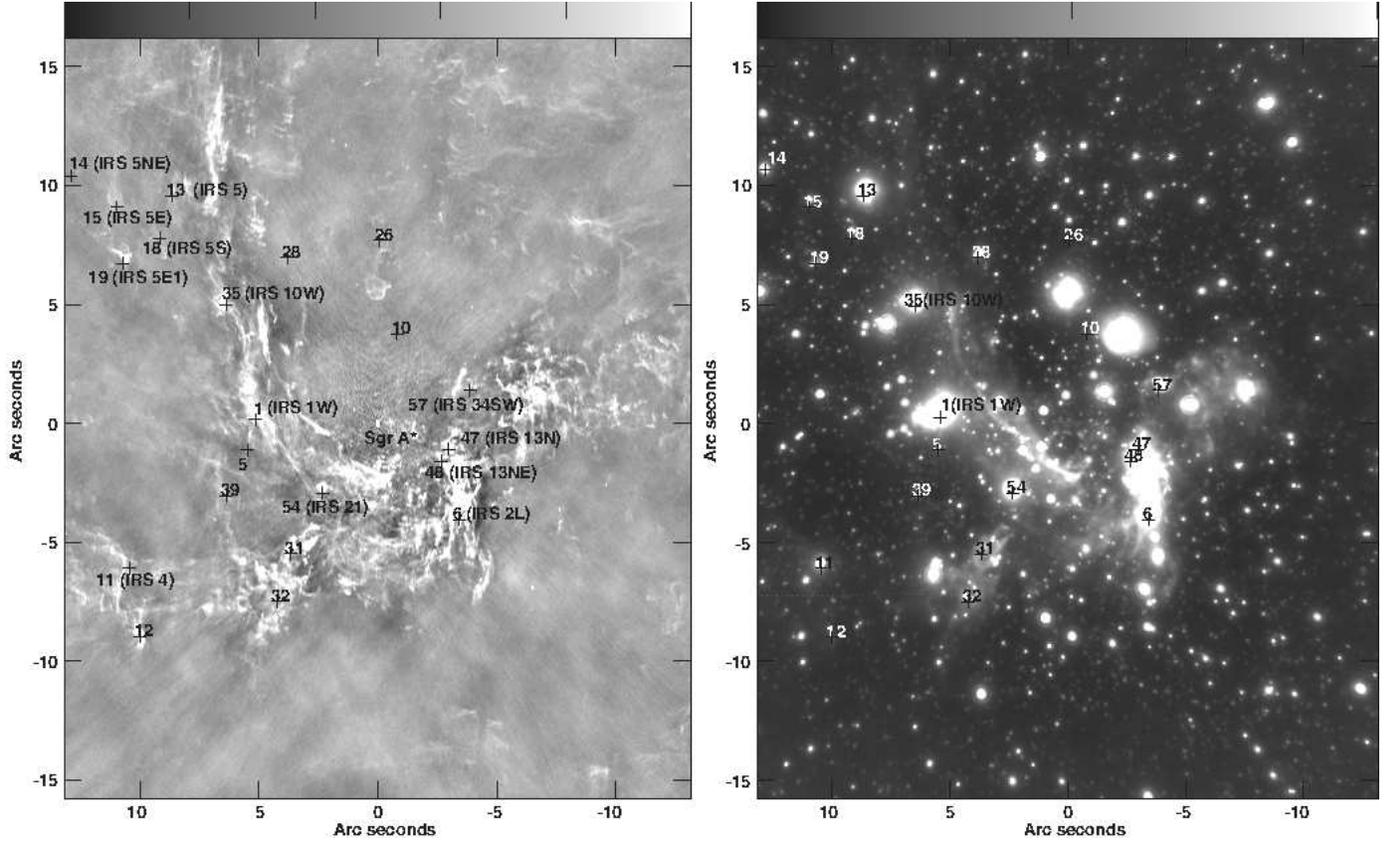}
\caption{
(\textit{a}) 
The distribution of YSO candidates, as listed in Table 2 
is superimposed on  a 34 GHz  continuum image based on a VLA A-array observation
with a resolution of $0.089''\times0.046''$ (PA$=-1.56^{\circ})$ (Yusef-Zadeh et al. 2015, in preparation). 
(\textit{b}) 
Same as (a) except an  L-band (3.8$\mu$m) image using the VLT. 
}
\end{figure}  
%\begin{figure}
%\ContinuedFloat
%\center
%\includegraphics[scale=0.9,angle=0]{fig4b_3cm_ysos_crop.eps}
%\end{figure}  

\begin{figure}
\figurenum{5}
\center
\includegraphics[scale=0.8,angle=0]{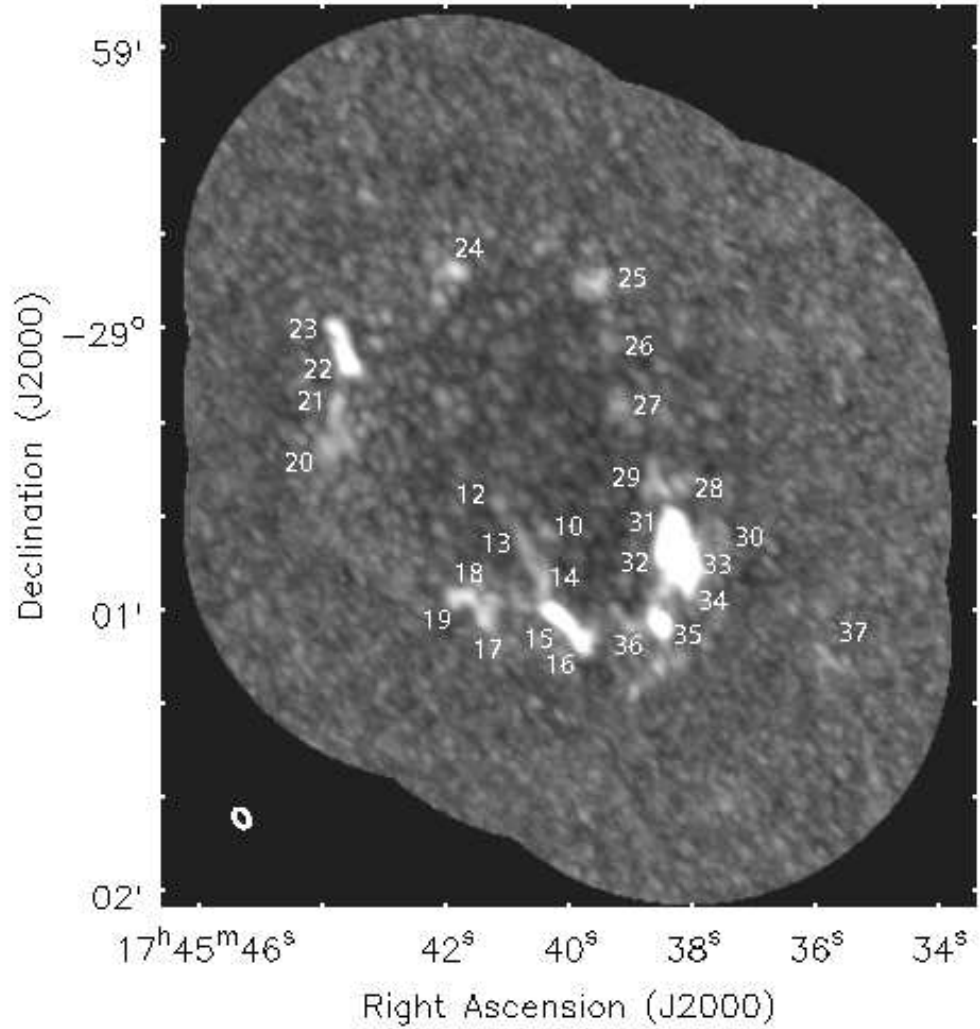}\\
\caption{
%{\it (a) Top }
A  map  of SiO (5-4) line emission from the molecular ring with 
a resolution of $3.6''\times2.4''$. Individual SiO (5-4) clumps, as listed in Table 4, 
are labeled. The SiO (5-4) data were originally published in 
Martin  et al. (2012). Clump F in the HCN map of the molecular ring (Christopher et al. 2005) coincides with 
SiO (5-4) clumps 23 and 24. 
}
\end{figure}  

\begin{figure}
\figurenum{6}
\center
\includegraphics[scale=0.9,angle=0]{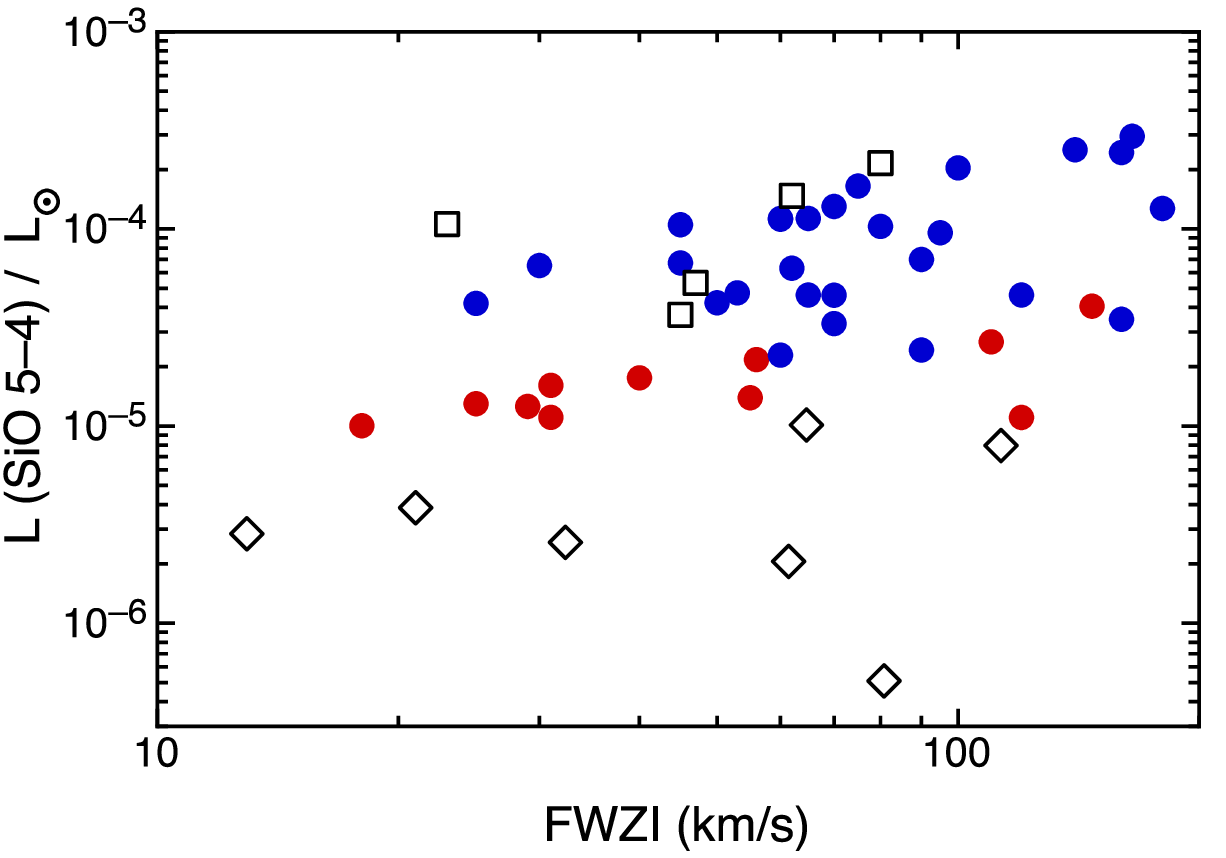}
\includegraphics[scale=0.9,angle=0]{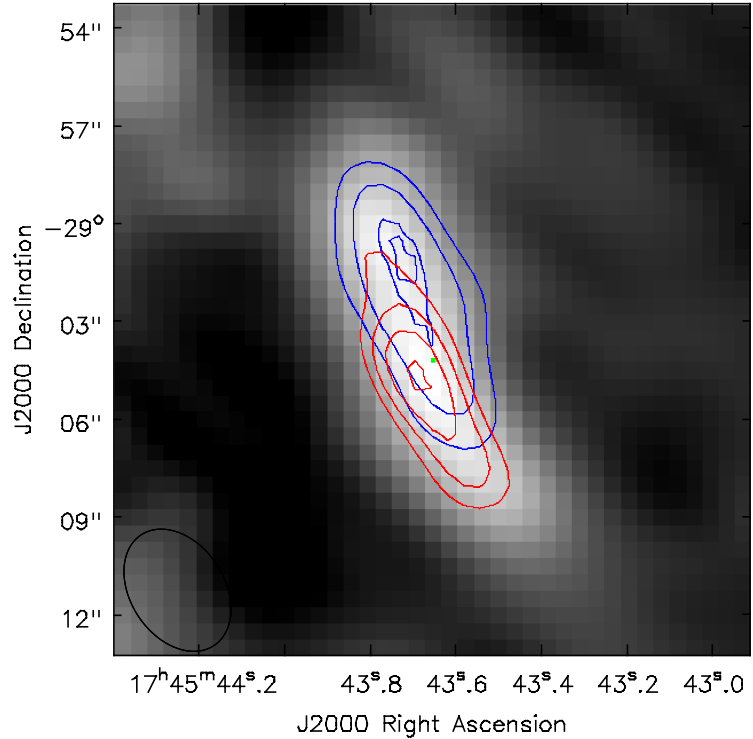}
\caption{
{\it (a) Top}
The SiO (5-4) luminosity vs the total linewidths FWZI of the line emission toward 11 clumps in the molecular ring (red)
based on ALMA observations,  
 and 26 clumps in the molecular ring based on this work (blue) 
as well as low and high-mass protostars in the Galactic disk (black diamonds and squares, respectively). 
{\it (b) Bottom}
Contours of SiO (5-4) line intensity between 30-55 \kms\, (blue) and 65-90 \kms\, (red) 
are superimposed on a grayscale image showing 
the  integrated SiO (5-4) line emission from  clumps 22 and 23 (Clump F in Christopher et al. 2005).
The ellipse at the bottom left corner represents the spatial resolution of SMA  data $3.6''\times2.4''$.  
}
\end{figure}  

\begin{figure}
\figurenum{7}
\center
\includegraphics[scale=0.7,angle=0]{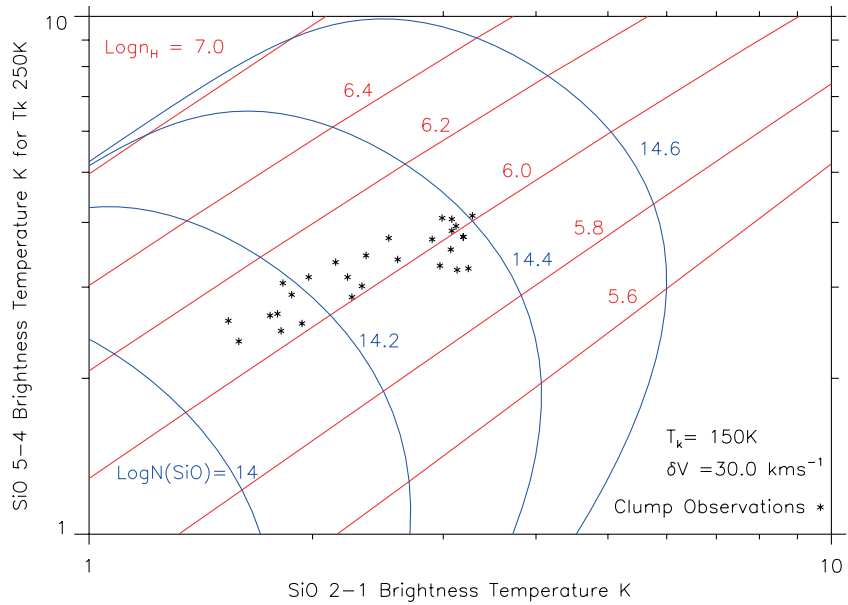}
\caption{
Intensities in the SiO (5-4) vs  SiO (2-1) lines  
for an assumed kinetic temperature of 150\,K
and FWHM ($\delta\,v$) of 30\,km\,s$^{-1}$.   Blue and red colors represent contours 
of constant SiO column density and gas density, respectively. 
The black star symbols  coincide with the observed brightness of  SiO (5-4) and (2-1)  clumps.
}
\end{figure}

\begin{figure}
\figurenum{8}
\center
\includegraphics[scale=0.65,angle=0]{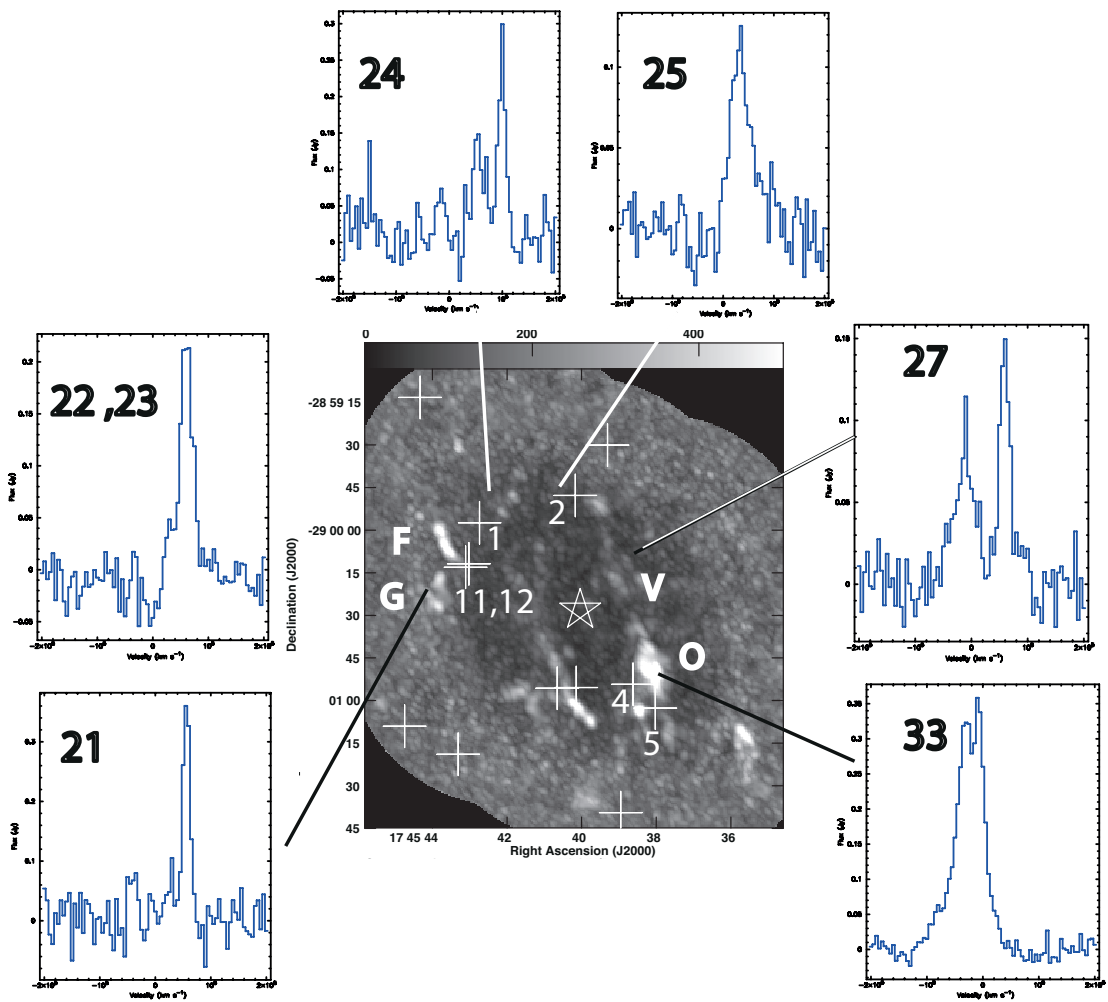}
\caption{
A grayscale image of  SiO (5-4) line emission             
with a  spatial resolution is $3.6''\times2.4''$. 
% (PA=1$^{\circ}$.2)
The spectra of  SiO (5-4) line emission from six  SiO (5-4) clumps 21, 22+23, 24, 25, 27, 33 (see Fig. 5) 
are displayed  as insets.
The crosses coincide with the location of  water masers. HCN (1-0) Clumps F, G, V and O 
are labeled (Christopher et al. 2005).
The location of Sgr A*  is shown
by an asterisk. The velocity range of individual plots is  between -2$\times10^5$ to 2$\times10^5$ \kms.}
%are  set at (1, 2,..., 7, 9, 12, 15, 18) Jy \kms\,  and are superimposed on a grayscale 
%5 GHz   continuum image (Yusef-Zadeh and Wardle 1993) with a spatial resolution of 0.67$''\times0.4''$ (PA=11$^{\circ}$).
\end{figure}

%vla data Nov 2004, a array AY152

\begin{figure}
\figurenum{9}
\center
\includegraphics[scale=0.6,angle=0]{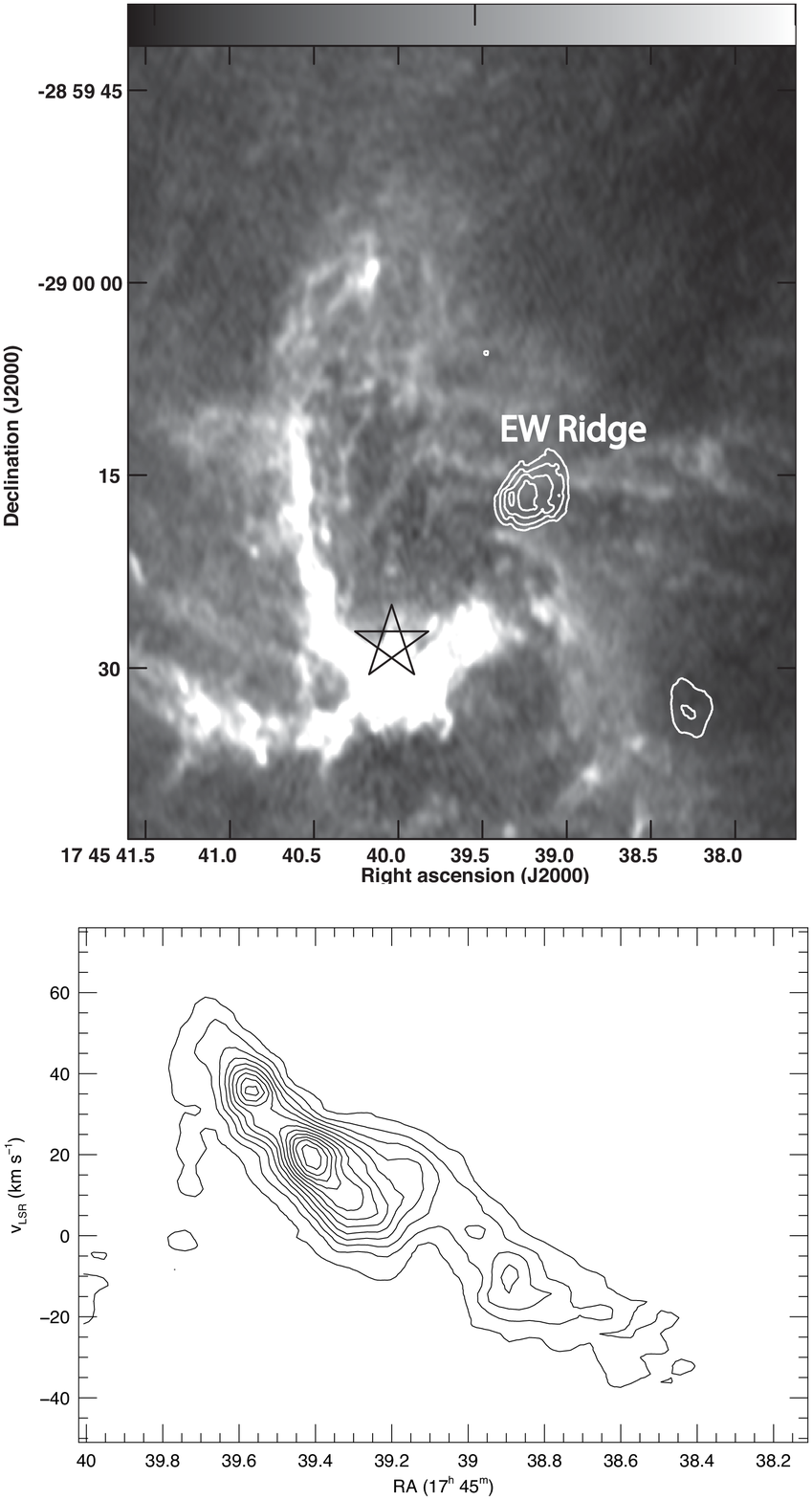}
\caption{
{\it (a Top)}
Contours of HCN(1-0) line emission from clump V
are set at 3, 3.4, 3.8, 4.2 Jy \kms\,  are superimposed on a 5 GHz  continuum image (Wardle and Yusef-Zadeh 1993).
A $20''$ horizontal structure EW Ridge with constant declination crosses contours of HCN emission from clump V 
and is noted on the grayscale 5 GHz image.  
The asterisk  symbol corresponds to the
location of Sgr A*. 
{\it (b Bottom)}
The position velocity diagram of [Ne\,{\sc ii}] line emission from the EW Ridge.
The cut runs along the brighter branch of the EW ridge, at PA$\approx-93^{\circ}$.  
%The position angle of the brighter branch of Ne\,{\sc ii} line emission is slightly different than 
%that of radio continuum branch. 
}
\end{figure}

%The contours are spaced quadratically, but 
%it would take me a while to figure out just what the contour levels are.  Could we leave that information out unless a referee asks for it 

\begin{figure}
\figurenum{10}
\center
\includegraphics[scale=0.25,angle=0]{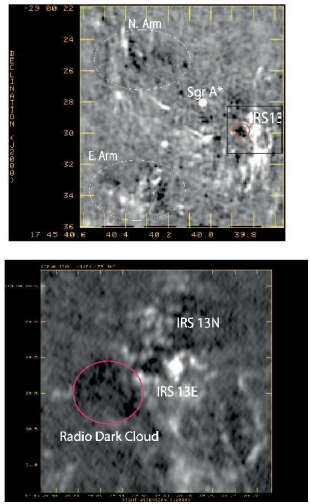}      
\center
\caption{
{\it{ a - Top}}
A  44 GHz  continuum image
with a spatial resolution of
$0.17''\times0.15''$ (PA=$-0.35^{\circ}$).
{\it{ b - Bottom}}
Enlarged view of the region drawn as a black square in (a) include both IRS 13E and IRS 13N clusters
The spatial  resolution of this image at 7mm is 82$\times42$\ milliarcsecond (PA=-5.5$^\circ$) (Yusef-Zadeh et al. 2014).
The white ellipses in (a) and the  red circle in (b) point to radio dark clouds. 
 }
\end{figure}

\end{document}